\documentclass[a4paper,10pt]{article}

\usepackage{latexsym}
\usepackage{amsmath,amsfonts,amssymb}
\usepackage{graphicx,color,epsfig}

\def\N{\mathbb{N}}

\def\R{\mathbb{R}}
\def\T{\mathbb{T}}
\def\Z{\mathbb{Z}}

\def\leq{\leqslant}
\def\geq{\geqslant}
\def\eps{\epsilon}

\newtheorem{Def}{Definition}[section]
\newtheorem{Thm}[Def]{Theorem}
\newtheorem{Lem}[Def]{Lemma}
\newtheorem{Pro}[Def]{Proposition}
\newtheorem{Cor}[Def]{Corollary}
\newtheorem{Claim}[Def]{Claim}

\begin{document}

\title{Wave generation in unidirectional chains of idealized neural oscillators}
\author{Bastien Fernandez$^{1}$\footnote{On leave from Centre de Physique Th\'{e}orique, CNRS - Aix-Marseille Universit\'e - Universit\'e de Toulon, Campus de Luminy, 13288 Marseille CEDEX 09 France}\, and Stanislav M.\ Mintchev$^2$}
\date{}   
\maketitle
\begin{center}
$^1$ Laboratoire de Probabilit\'es et Mod\`eles Al\'eatoires\\
CNRS - Universit\'e Paris 7 Denis Diderot\\
75205 Paris CEDEX 13 France\\
{\tt fernandez@math.univ-paris-diderot.fr}\\
\medskip
$^2$ Department of Mathematics\\The Cooper Union\\
New York NY 10003 USA\\
{\tt mintchev@cooper.edu}\\
\end{center}

\begin{abstract}
We investigate the dynamics of unidirectional semi-infinite chains of type-I oscillators that are periodically forced at their root node, as an archetype of wave generation in neural networks. In previous studies, numerical simulations based on uniform forcing have revealed that trajectories approach a traveling wave in the far-downstream, large time limit. While this phenomenon seems typical, it is hardly anticipated because the system does not exhibit any of the crucial properties employed in available proofs of existence of traveling waves in lattice dynamical systems. Here, we give a full mathematical proof of generation under uniform forcing in a simple piecewise affine setting for which the dynamics can be solved explicitly. In particular, our analysis proves existence, global stability, and robustness with respect to perturbations of the forcing, of families of waves with arbitrary period/wave number in some range, for every value of the parameters in the system.

\end{abstract}

\leftline{\small\today.}

\section{Introduction}
Signal propagation in the form of waves is a ubiquitous feature of the functioning of neural networks. Waves transmitting electrical activity across neural structures have been observed in a large variety of situations, both in artificially grown cultures and in living brain tissues, see {\sl e.g}\ \cite{KDetal94} and \cite{JM07} for an instance of each case; many other examples can be found in the literature.

This experimental phenomenology has fostered numerous computational and analytical studies on theoretical models for wave propagation. To mention a single category of exact results, one can cite proofs of existence of waves with context dependent shape: fronts, pulses, periodic wave trains, etc, both in full voltage/conductance models and in firing rate models, see {\sl e.g.}\ \cite{CB03,EM93,GE02}. In parallel, numerical studies have investigated propagation features such as firing synchrony within cortical layers, and their dependence on dynamical ingredients: feedback, surrounding noise, or external stimulus, see for instance \cite{DGA99,JMY13,LSetal03,SK95}.

Our paper aims to develop a rigorous mathematical investigation of how the global dynamics of a (simple model of a) neural network may cause it to organize to a wave behavior, in spite of being forced by a rather unrelated signal. Given that the natural setting of neural ensembles typically features an external environment that is prone to providing an array of irregular stimuli, it seems that such forcing is in no way {\sl a priori} tailored toward generating periodic patterns in layered ensembles. Nevertheless, recordings from tissues suggest that self-organization will often ensue despite this obvious and inherent mismatch. 

In order to get insight into the generation of periodic traveling waves through {\sl ad hoc} stimulus, we consider unidirectional chains of coupled oscillators. Inspired by the propagation of synfire waves through cortical layers \cite{A82}, such systems can be regarded as basic phase variable models of feed-forward networks featuring synchronized groups, in which each pool can be treated as a phase oscillator that repetitively alternates a refractory period with a firing burst. In addition, chains with unidirectional coupling as in equation \eqref{FEEDFORWARD} below are representative of some physiological systems, such as central pattern generators \cite{DCBCB99}. Also, acyclic chains of type-I oscillators have been used as simple examples for the analysis of network reliability \cite{lin_esbrown_lsyoung:2007}.

\begin{figure}[h]
\centerline{\includegraphics{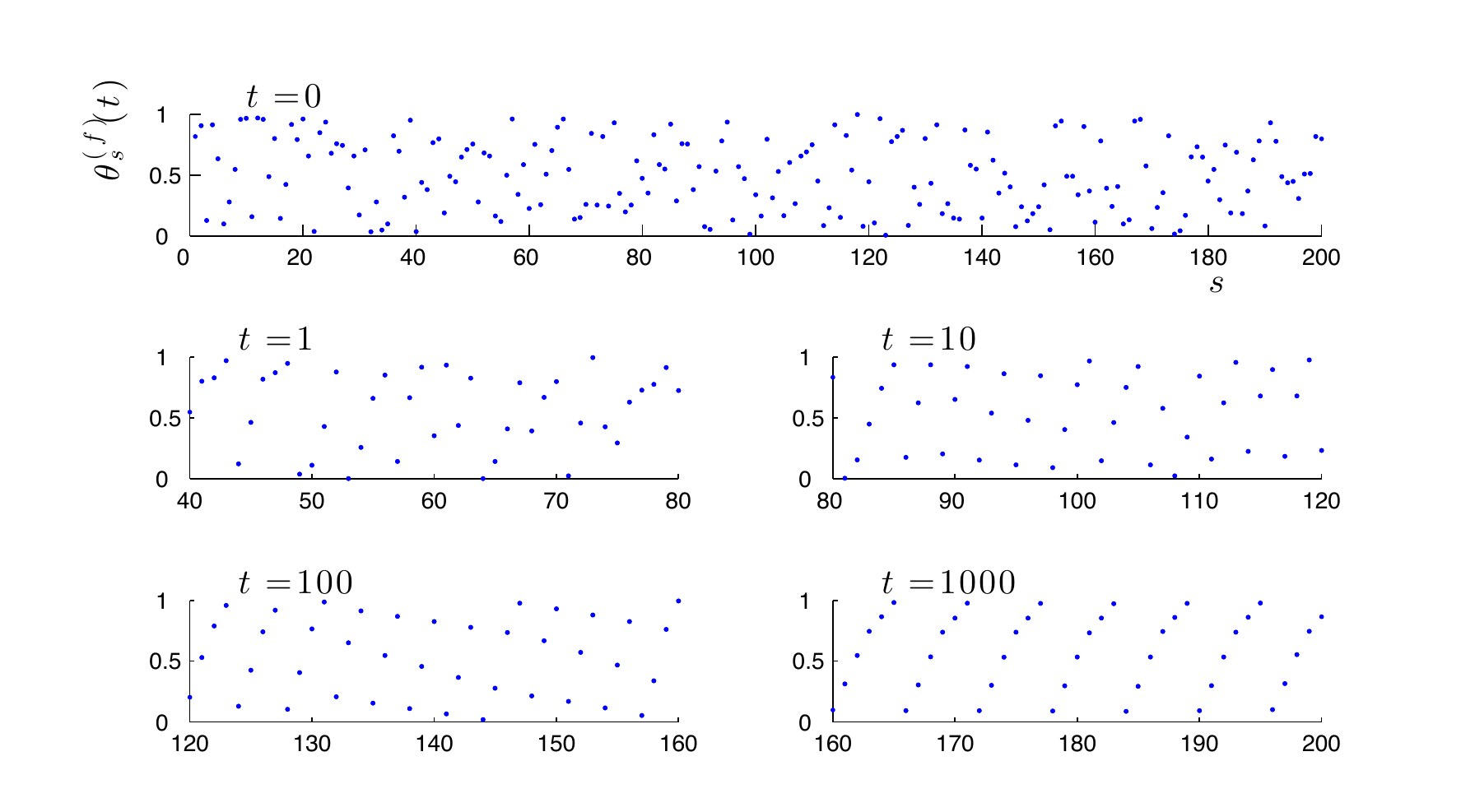}}
\caption{Snapshots ($t=0$, $1$, $10$, $100$, $1000$) of oscillator phases $\theta_s^{(f)}(t) \in [0,1)$ with periodic boundary condition ($s=1,\ldots,200$) for a typical trajectory issued from a random initial condition, when forcing with a uniform signal $f(t)=t/\tau$. Clearly, a traveling wave with periodic shape emerges in the far-downstream large time limit (i.e., as $s$ and $t \to \infty$).}
\label{SNAPSHOTS}
\end{figure}

More specifically, the model under consideration deals with chains of coupled phase oscillators whose dynamics is given by the following coupled ODEs
\begin{equation}
\frac{d\theta_s}{dt}=\omega+\epsilon\Delta(\theta_s)\delta(\theta_{s-1}),\ \forall s=1,2,\cdots 
\label{FEEDFORWARD}
\end{equation}
where $\theta_s\in\T:=\R/\Z$ and 
\begin{itemize}
\item[$\bullet$] $\epsilon,\omega\in\R^+$. Up to a rescaling of time, we can always assume that $\omega=1$.
\item[$\bullet$] $\Delta$ is the so-called phase response curve (PRC). Recall that a type-I oscillator is one for which the PRC is a non-negative one-humped function \cite{Hans_Mato_Meun:1995neur_comp}.
\item[$\bullet$] $\delta$ mimics incoming stimuli from the preceding node and also takes the form of a unimodal function. 
\item[$\bullet$] the first oscillator at site $s=0$ evolves according to some forcing signal $f$ (external stimulus), {\sl i.e.}\ we have $\theta_0(t)=f(t)$ for all $t\in\R^+$. The forcing is 
assumed to be continuous, increasing and periodic ({\sl viz.}\ there exists $\tau\in\R^+$ such that $f(t+\tau)=f(t)+1$ for all $t\in\R^+$). 
\end{itemize}  
Numerical simulations of the system \eqref{FEEDFORWARD} for some smooth functions $\Delta$ and $\delta$ and forcing signals such as the uniform function $f(t)=t/\tau$, have revealed that the asymptotic dynamics settles to a highly-organized regime, independently of initial conditions \cite{smmintchev_lsyoung:2009}. As $t \to +\infty$, the phase $\theta_s(t)$ approaches the perturbation of a periodic function - the same function $f_0$ ($f_0 \neq f$) at each site, up to an appropriate time shift $\alpha$ - and this perturbation is attenuated by the chain in going further and further down. In brief terms, traveling waves are typically observed in the far-downstream, large time limit; see Fig.~\ref{SNAPSHOTS}. Mathematically speaking, this means that, letting $\{ \theta_s^{(f)}(t) \}_{s \in \N}$ denote the solution to \eqref{FEEDFORWARD} with forcing $f$ (and say, typical initial condition), there exists a periodic function $f_0$ and a time shift $\alpha\in\R^+$ such that we have
\[
\lim_{s \to +\infty} \left( \limsup_{t \to +\infty} \left| \theta_s^{(f)} (t) - f_0(t + \alpha s) \right| \right) = 0\ (\text{\rm mod}\ 1).
\]
Moreover, this phenomenon occurs for arbitrary forcing period\footnote{The wave period is neither necessarily equal to $\tau$, nor to $1/\omega$; see~\cite{LM14} for an illustration.} in some range and also appears to be robust to changes in the coupling intensity $\epsilon$.

This behavior was somehow unanticipated because \eqref{FEEDFORWARD} does not reveal any crucial properties usually required in the proofs of wave existence, such as the monotonicity of the profile dynamics (analogous property to the maximum principle in parabolic PDEs), see {\sl e.g.}\ \cite{BCC03,M-P97,ZHH93} for lattice differential equations and \cite{CF04a,L89,W02} for discrete time recursions.\footnote{In the end of section \ref{S-BASIC}, we show that monotonicity with respect to 'pointwise' ordering on sequences in $\R^{\Z^+}$ fails in this system.} In this context, proving the existence of waves remains unsolved and so is the stability problem, not to mention any justification of the generation phenomenon when forcing with {\sl ad hoc} signal. Notice however that, by assuming the existence of waves and their local stability for the single-site dynamics, a proof of stability for the whole chain has been obtained and applied to the design of numerical algorithms for the double-precision construction of wave shapes \cite{LM14}. (Our stability proof here is inspired by this one.)

In order to get mathematical insights into wave generation under {\sl ad hoc} forcing, here, we analyze simple piecewise affine systems for which the dynamics can be solved explicitly. This analysis can be viewed as an exploratory step in the endeavor of searching for full proofs in (more general) nonlinear systems. Hence, the functions $\Delta$ and $\delta$ are both assumed to be piecewise constant on the circle, taking on only the two distinct values of 1 ('on') or 0 ('off'). 

In this setting, our analysis shows that the numerical phenomenology can be mathematically confirmed. For all parameter values, we prove the existence of TW with arbitrary period in some interval, and their global stability with respect to initial perturbations in the phase space $\T^\N$, not only when the forcing at $s=0$ is chosen to be a TW shape but also for an open set of periodic signals with identical period. In addition, this open set is shown to contain uniform forcing $f(t)=t/\tau$ provided that the coupling intensity $\epsilon$ is sufficiently small. 

The paper is organized as follows. The next section contains the accurate definition of the initial value problem, the basic properties of the associated flow and the statements of the main results. The rest of the paper is devoted to proofs. In section \ref{S-EXIST}, we prove the existence of TW by establishing an explicit expression of their shape. We study the TW stability with respect to initial conditions in section \ref{S-STAB} by considering the associated stroboscopic dynamics, firstly for the first site, and then for the second site, from where the stability of the full chains is deduced. Finally, stability with respect to changes in forcing is shown in section \ref{S-GENER}, as a by-product of the arguments developed in the previous sections. Section \ref{S-CONCL} offers some concluding remarks.

\section{Definitions, basic properties, and main results}
As mentioned before, the dynamical systems under consideration in this paper are special cases of equation \eqref{FEEDFORWARD} in which the stimulus $\delta$ and the PRC $\Delta$ are (non-negative and normalized) square functions, namely\footnote{{\bf Notations.} $\Z^+=\{0,1,2,\cdots\}$, $\N=\{1,2,\cdots\}$ and $\lfloor\cdot\rfloor$ denotes the floor function (recall that $0\leq x-\lfloor x\rfloor<1$ for all $x\in\R$). $\vartheta$ and $\vartheta_s$ denote elements in $\T$ whereas $\theta$ and $\theta_s$ represent functions of the real positive variable with values in $\T$.}
\[
\delta(\vartheta)=\left\{\begin{array}{l}
1\ \text{if}\ \vartheta\ \text{mod}\ 1\in [0,a_0]\\
0\ \text{otherwise}
\end{array}\right.
\quad\text{and}\quad
\Delta(\vartheta)=\left\{\begin{array}{l}
1\ \text{if}\ \vartheta\ \text{mod}\ 1\in [0,a_1]\\
0\ \text{otherwise}
\end{array}\right.\quad\forall \vartheta\in\T.
\]
(Of note, the stimulus $\delta$ can be made arbitrarily brief by choosing $a_0$ arbitrarily close to 0. Moreover, that the two intervals $[0,a_0]$ and $[0,a_1]$ have the same left boundary is a simplifying assumption that reduces the number of parameters. Other cases are of interest, for instance, oscillators hearing poorly when transmitting, which corresponds to non-overlapping intervals.)

More formally, we shall examine the following system of coupled differential equations for semi-infinite configurations $\{\theta_s\}_{s\in\Z^+}\in\T^{\Z^+}$
\begin{align}
&\frac{d\theta_s}{dt}=\left\{\begin{array}{cl}
1+\epsilon&\text{if}\ \theta_{s-1}-\lfloor \theta_{s-1}\rfloor\leq a_0\ \text{and}\ \theta_{s}-\lfloor \theta_{s}\rfloor\leq a_1\\
1&\ \text{otherwise}
\end{array}\right.\ \forall s\in\N,t>0
\label{DEFDYNAM}\\
&\theta_0(t)=f(t),\ \forall t\in \R^+\hfill\nonumber\\
&\theta_s(0)=\vartheta_s,\ \forall s\in\N,\hfill\nonumber
\end{align}
where 
\begin{itemize}
\item[-] the forcing signal $f$ is assumed to be a Lipschitz-continuous, $\tau$-periodic\footnote{We refer to the $\T$-valued $f$ as $\tau$-periodic provided that $f(t+\tau)=f(t)+1$ for all $t\in\R^+$.} ($\tau>0$), and increasing function with slope (wherever defined) at least 1,\footnote{In order to ensure continuous dependence of solutions of equation \eqref{DEFDYNAM} on forcing, we actually only need  that the forcing slope be bounded below by a positive number, the same number for all forcing. The choice 1 for the bound here is consistent with the minimal rate at which solutions can grow. Typically, $f$ is thought of being piecewise affine or even simply affine.} and satisfying $f(0)=0$,
\item[-] $\{\vartheta_s\}_{s\in\N}\in\T^{\N}$ is an arbitrary initial configuration,
\item[-] $\epsilon>0$ and $a_0,a_1\in (0,1)$ are arbitrary parameters.
\end{itemize}

Solutions of equation \eqref{DEFDYNAM} are in general denoted by $\{\theta_s^{(f)}(t)\}_{s\in\N}$ but the notation $\{\theta_s^{(f)}(\vartheta_1,\cdots,\vartheta_s,t)\}_{s\in\N}$ is also employed when the dependence on initial condition needs to be explicitly mentioned. 

\subsection{Basic properties}\label{S-BASIC}
The solutions of equation \eqref{DEFDYNAM} have a series of basic properties which we present and succinctly argue in a rather informal way. These facts can be formally established by explicitly solving the dynamics. The details are left to the reader.

\noindent
{\bf Existence of the flow.} Given any forcing signal $f$ and any initial condition $\{\vartheta_s\}_{s\in\N}$, for every $s\in\N$, there exists a unique function $t\mapsto \theta_s^{(f)}(t)$ which satisfies equation \eqref{DEFDYNAM} for all $t\in\R^+$. This function $\theta_s^{(f)}$ is continuous, increasing, and piecewise affine with alternating slope in $\{1,1+\epsilon\}$. Moreover each piece of slope 1 must have length $\geq 1-a_1$.

These facts readily follow from solving the dynamics inductively down the chain. Assuming that $t\mapsto \theta_{s}^{(f)}(t)$ is given for some $s\in\N$ (or considering the forcing term $f$ if $s=0$), the slope of the first piece of $t\mapsto\theta_{s+1}^{(f)}(t)$ only depends on the relative position of $\vartheta_{s+1}$ with respect to $a_1\ (\text{\rm mod}\ 1)$ and of $\vartheta_s$ with respect to  $a_0\ (\text{\rm mod}\ 1)$. The length of this piece depends on its slope, on $\vartheta_{s+1}$ and on the smallest $t>0$ such that $\theta_{s}(t)\in\{0,a_0\}\ (\text{\rm mod}\ 1)$; this infimum time has to be positive. By induction, this process generates the whole function $\theta_{s+1}^{(f)}$ by using the location of $\theta_s$ and $\theta_{s-1}$ at the end of each piece, and the next time when $\theta_s\in \{0,a_0\}\ (\text{\rm mod}\ 1)$.

\noindent
{\bf Continuous dependence on inputs.} Endow $\T^{\Z^+}$ with pointwise topology and, given $T>0$, endow continuous and monotonic functions of $t\in [0,T]$ with uniform topology and norm $\|\cdot\|$. For every $s\in\N$ and $T>0$, the quantity $\|\theta_s^{(f)}|_{[0,T]}\|$ continuously depends both on the forcing signal $f|_{[0,T]}$ and on the initial condition $\{\vartheta_s\}_{s\in\N}$. 

Indeed, if two forcing signals $f$ and $g$ are close, then the lower bound on their derivatives implies that the respective times in $[0,T]$ at which they reach $\{0,a_0\}\ (\text{\rm mod}\ 1)$ must be close. If, in addition, the initial conditions $\vartheta_1$ and $\xi_1$ are close, then the trajectories $t\mapsto \theta_1^{(f)}(\vartheta_1,t)$ and $t\mapsto \theta_1^{(g)}(\xi_1,t)$ alternate their slopes at close times; hence $\|\theta_1^{(f)}(\vartheta_1,\cdot)|_{[0,T]}-\theta_1^{(g)}(\xi_1,\cdot)|_{[0,T]}\|$ must be small. Since the slopes are at least 1, the respective times at which $\theta_1^{(f)}(\vartheta_1,\cdot)|_{[0,T]}$ and $\theta_1^{(g)}(\xi_1,\cdot)|_{[0,T]}$ reach $\{0,a_0\}\ (\text{\rm mod}\ 1)$ are close. By repeating the argument, we conclude that $\|\theta_2^{(f)}(\vartheta_1,\vartheta_2,\cdot)|_{[0,T]}-\theta_2^{(g)}(\xi_1,\xi_2,\cdot)|_{[0,T]}\|$ must be small when $\vartheta_2$ and $\xi_2$ are sufficiently close. Then, the result for an arbitrary $s\in\N$ follows by induction.

\noindent
{\bf Semi-group property.} As suggested above, a large part of the analysis consists in focusing on the one-dimensional dynamics of the first oscillator $\theta_1$ forced by the stimulus $f$, prior to extending the results to subsequent sites. Indeed, the dynamics of the oscillator $s$ can be regarded as a forced system with input signal $\theta_{s-1}$.

For the one-dimensional forced system, letting $R^T\theta(t)=\theta(t+T)$ for all $t\in\R^+$ denotes the time translations, we shall especially rely on the 'semi-group' property of the flow, {\sl viz.}\ if $\theta_1^{(f)}(\vartheta_1,\cdot)$ is a solution with initial condition $\vartheta_1$ and forcing $f$, then, for every $T\in\R^+$, $R^T\theta_1^{(R^Tf)}(\theta_1^{(f)}(\vartheta_1,T),\cdot)$ is a solution with initial condition $\theta_1^{(f)}(\vartheta_1,T)$ and forcing $R^Tf$.

\noindent
{\bf Monotonicity failure.} Consider the following partial order on sequences in $\R^{\N}$ (employed in typical proofs of existence of TW in lattice dynamical systems). We say that $\{\vartheta_s\}_{s\in\N}\leq \{\xi_s\}_{s\in\N}$ iff $\vartheta_s\leq \xi_s$ for all $s$. Clearly, we may have 
\[
\{\theta_s^{(f)}\left(\vartheta_1,\cdots,\vartheta_s,t\right)\}_{s\in\N}\leq \{\theta_s^{(f)}(\xi_1,\cdots,\xi_s,t)\}_{s\in\N},
\]
for some $t\in\R^+$ and yet 
\[
\{\theta_s^{(f)}(\vartheta_1,\cdots,\vartheta_s,t')\}_{s\in\N}\not\leq \{\theta_s^{(f)}(\xi_1,\cdots,\xi_s,t')\}_{s\in\N}
\]
for another $t'>t$. For instance, it suffices to choose $\vartheta_1\in (0,a_0),\xi_1\in (a_0,1)$ and $\vartheta_2\leq \xi_2\in (0,a_1)$ with $\vartheta_2$ and $\xi_2$ sufficiently close. See Figure \ref{MONOTONICITY_FAIL} below.

\begin{figure}[h]
\centerline{\includegraphics[scale=0.4]{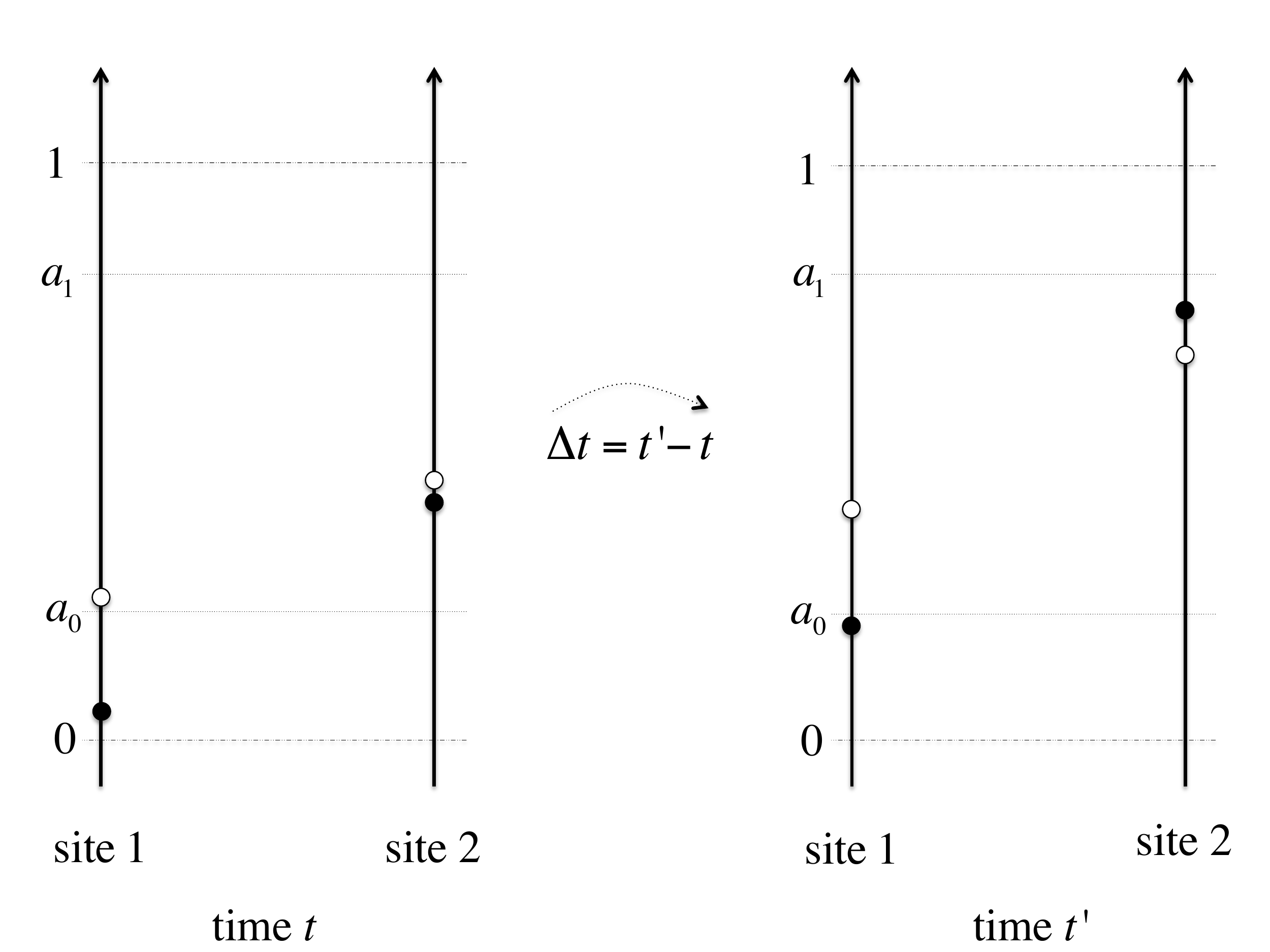}}
\caption{Illustration of non-monotonicity of the profile dynamics. The solid and hollow black dots respectively denote $\theta_1^{(f)}(\vartheta_1,\cdot )$ and $\theta_1^{(f)}(\xi_1, \cdot )$ at site $s=1$, and $\theta_2^{(f)}(\vartheta_1,\vartheta_2,\cdot )$ and $\theta_2^{(f)}(\xi_1, \xi_2,\cdot )$ at site $s=2$.}
\label{MONOTONICITY_FAIL}
\end{figure}

\subsection{Main results}
\begin{figure}[h]
\centerline{\includegraphics{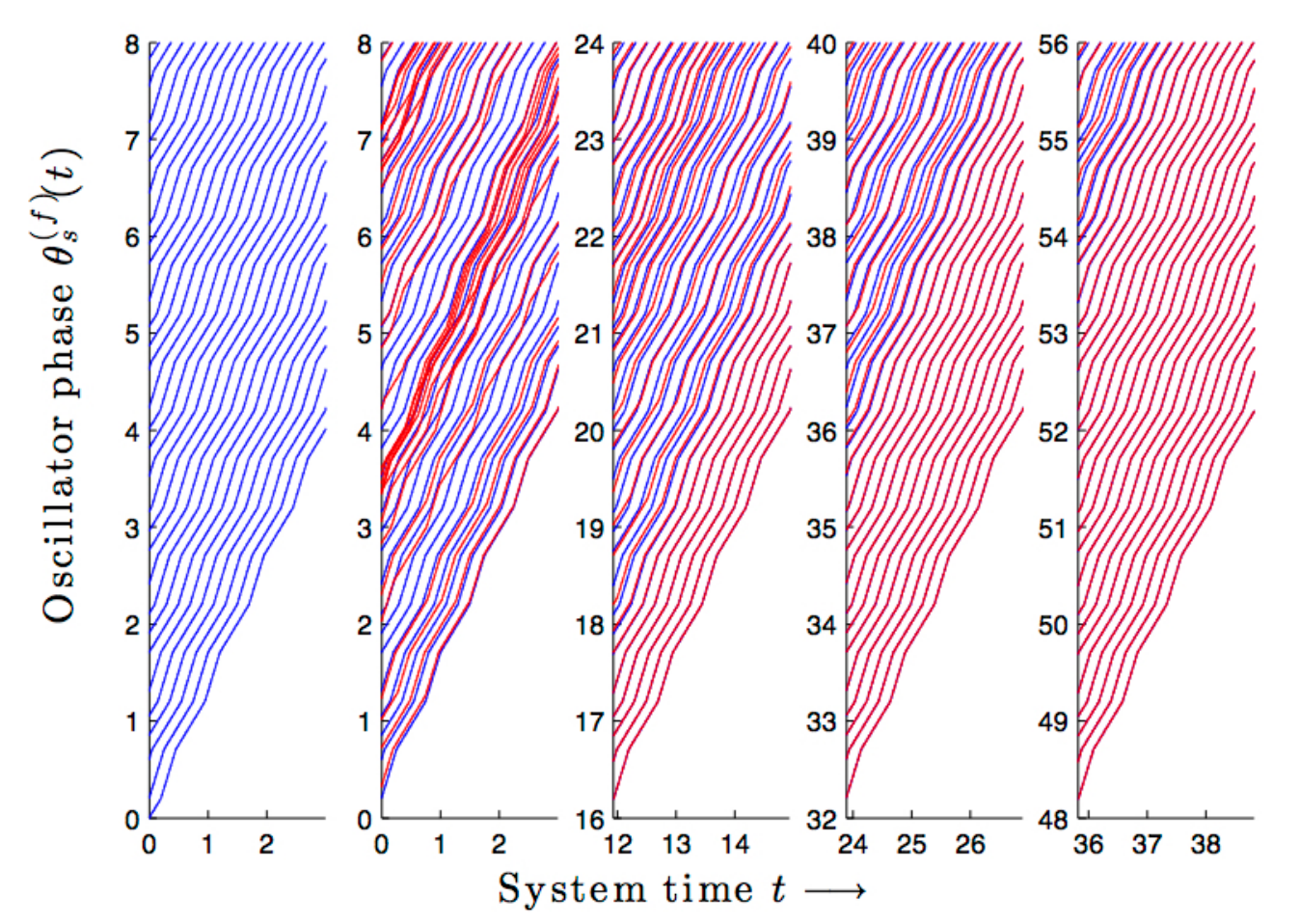}
}
\caption{{\bf Left panel:} Traveling wave solution $\theta_s^{(f)}(t)=f(t+\alpha s)$ for $\alpha = 0.2$ and $s =1,\ldots,30$ vs.~time $t$, together with the forcing signal $f$ (which satisfies condition (C) in Thm~\ref{MAINRES1}, so as to ensure that the wave is stable -- see text). {\bf Subsequent panels:} Example of a trajectory $\theta_s^{(f)}(t)$ for $s=1,\ldots,30$ and disjoint consecutive time windows. The initial phases have been chosen randomly and the forcing signal $f$ is the same as in the left panel. Clearly, the convergence $\theta_s^{(f)}(t) \to f(t+\alpha s)$ occurs at the first site s=1 and then propagates down the chain, as described in the text; the higher the value of $s$ is, the longer the transient time $\theta_s^{(f)}(t)$ needs in order to get close to $f(t+\alpha s)$.}
\label{TW_REP-AND-CONVERGE}
\end{figure}

A solution $\{\theta_s^{(f)}(t)\}_{s\in \N,t\in\R^+}$ of the system \eqref{DEFDYNAM} is called a {\bf traveling wave} (TW) if there exists $\alpha\in\R^+$ such that 
\[
\theta_{s}^{(f)}(t)=f(t+\alpha s), \forall s\in\N,t\in\R^+.
\]
In other words, a TW is a solution for which the forcing signal $f$ exactly repeats at every site, modulo an appropriate time shift: the phase at any given site $s$ and time $t$ mimics the phase at the previous site and time $t-\alpha$ (see Fig.~\ref{TW_REP-AND-CONVERGE}, left panel).  For such solutions, the forcing signal also plays the role of the wave shape and the quantity $\frac{\alpha}{\tau}$ represents the wave number. In our setting, any TW shape/forcing signal obviously has to be a piecewise affine function with slopes only taking the values $1$ and $1+\epsilon$.

As we shall see below, TW exist that are asymptotically stable, not only with respect to perturbations of the initial phases $\{\vartheta_s\}$, but more importantly, also with respect to changes in the forcing signal. For convenience, we first separately state existence and uniqueness.
\begin{Thm}
(Existence.) For every $\epsilon,a_0$ and $a_1$, there exists a (non-empty) interval $I_{\epsilon,a_0,a_1}$ and for every $\tau\in I_{\epsilon,a_0,a_1}$, there exist a $\tau$-periodic forcing signal $f$ and $\alpha\in (0,\tau)$ such that $\{f(t+\alpha s)\}_{s\in \N,t\in\R^+}$ is a TW. 

(Uniqueness.) For every $\tau\in I_{\epsilon,a_0,a_1}$, the forcing signal $f$ and the shift $\alpha$ as above are unique, provided that the following constraint is required

{\rm (C)} $f$ is a piecewise affine forcing signal whose restriction $f|_{(0,\tau)}$ has slope $1+\epsilon$ only on a sub-interval whose left boundary is $\alpha$.
\label{MAINRES1}
\end{Thm}
For the proof, see section \ref{S-EXIST}, in particular Corollary \ref{EXISTW}. Notice that constraint (C) has no intrinsic interest other than unambiguously identifying appropriate TW shapes for the stability statement. To identify stable waves matters because the system \eqref{DEFDYNAM} also possesses neutral and unstable traveling waves.

For stability, we shall use notions that are appropriate to forced systems, and adapted to our setting. In particular, since the information flow is unidirectional here, it is natural to only require that perturbations relax in pointwise topology, rather than in uniform topology. Therefore, we shall consider the dynamics on arbitrary finite collections of sites which, without loss of generality, can be chosen to be the first $s$ sites, for an arbitrary $s$. Moreover, there is no need for local stability considerations here because we shall be concerned with TW for which the basin of attraction is as large as it can get from a topological viewpoint.

Accordingly, we shall say that a TW $(f,\alpha)$ is {\bf globally asymptotically stable} if there exists a sequence $\{\vartheta_{\text{unst},s}\}_{s\in\N}\in\T^\N$ such that, for every $s\in\N$ 
and every initial condition for which $\vartheta_{r}\neq \vartheta_{\text{unst},r}$ for all $r\in \{1,\cdots,s\}$, we have 
\[
\lim_{t\to+\infty}\theta_{r}^{(f)}(\vartheta_1,\cdots,\vartheta_r,t)-f(t+\alpha r)=0\ (\text{\rm mod}\ 1),\ \forall r\in\{1,\cdots,s\}.
\]
In other words, a solution is globally asymptotically stable if its basin of attraction is as large as it can get from a topological viewpoint. Again, the exceptional initial conditions $\vartheta_{\text{unst},s}$ must exist for fixed-point index reasons. As can be expected, our next statement claims global asymptotic stability of waves, provided they are suitably chosen according to the above criterion.
\begin{Thm}
There exists a (non-empty) sub-interval $I'_{\epsilon,a_0,a_1}\subset I_{\epsilon,a_0,a_1}$ such that for every $\tau\in I'_{\epsilon,a_0,a_1}$, the TW $(f,\alpha)$ determined by constraint {\rm (C)} is globally asymptotically stable. 
\label{MAINRES15}
\end{Thm}
See Fig.~\ref{TW_REP-AND-CONVERGE} for an illustration of this result. Theorem \ref{MAINRES15} is proved in section \ref{S-STAB}, see especially the concluding statement Corollary \ref{STABTW}. In addition, for initial conditions not satisfying the stability condition, our proof shows that the first coordinate $s$ for which this condition fails asymptotically approaches an unstable periodic solution. 

In Theorem \ref{MAINRES15}, we claim stability of the TW $(f,\alpha)$ with respect to perturbations of initial conditions when forcing with $f$. We now consider the analogous property when the forcing signal is also perturbed. As described in the introduction, in general, for any given site $s$, one may not expect relaxation to $f(t+\alpha s)$ exactly but rather to a perturbation of this signal, which is itself attenuated as $s \to \infty$. Accordingly, we shall say that a TW $(f,\alpha)$ is {\bf robust with respect to perturbations of the forcing} if, for every forcing signal $g$ in a $C^1$-neighborhood of $f$, there exists a neighborhood $V$ (product topology) of $\{f(\alpha s)\}_{s\in\N}$ such that, for every $\{\vartheta_s\}_{s\in\N}\in V$, we have 
\[
\lim_{s\to +\infty}\limsup_{t\to +\infty}|\theta_s^{(g)}(\vartheta_1,\cdots,\vartheta_s,t)-f(t+\alpha s)|=0\ (\text{\rm mod}\ 1),
\]
as indicated in the Introduction. 
Robustness with respect to forcing perturbations is expected to hold in general smooth systems of the form \eqref{FEEDFORWARD} (as is the global stability of TW). In the current piecewise affine setting, the phase $\theta_s^{(g)}(t)$ actually relaxes to $f(t+\alpha s)$ at every site $s$ (a stronger result), as described in the following statement.
\begin{Thm}
For every $\tau\in I'_{\epsilon,a_0,a_1}$, there exists a $C^1$-neighborhood $U$ of $f$ such that, for every $g\in U$, the TW $(f,\alpha)$ determined by constraint {\rm (C)} globally attracts solutions of the system \eqref{DEFDYNAM} with forcing $g$. That is to say, there exists a sequence $\{\vartheta_{\text{unst},s}^{(g)}\}_{s\in\N}\in\T^\N$ such that for every $s\in\N$ and every initial condition for which $\vartheta_{r}\neq \vartheta_{\text{unst},r}^{(g)}$ for all $r\in \{1,\cdots,s\}$, we have 
\[
\lim_{t\to+\infty}\theta_{r}^{(g)}(\vartheta_1,\cdots,\vartheta_r,t)-f(t+\alpha r)=0\ (\text{\rm mod}\ 1),\ \forall r\in\{1,\cdots,s\}.
\]

In addition, for $\epsilon$ small enough (depending on $a_0,a_1$), there exists a (non-empty) sub-interval $I''_{\epsilon,a_0,a_1}\subset I'_{\epsilon,a_0,a_1}$ such that, for every $\tau\in I''_{\epsilon,a_0,a_1}$, the neighborhood $U$ contains the uniform forcing $g(t)=\frac{t}{\tau}$, for all $t\in\R^+$.
\label{MAINRES2}
\end{Thm}
Theorem \ref{MAINRES2} is established in section \ref{S-GENER} and implies in particular robustness with respect to forcing perturbations.
\begin{Cor}
For every $\tau\in I'_{\epsilon,a_0,a_1}$, the TW determined by the constraint {\rm (C)} is robust with respect to perturbations of the forcing.
\end{Cor}

\section{Existence of traveling wave solutions}\label{S-EXIST}
An alternative characterization of TW solutions can be given as follows
\[
\theta_{s+1}(t)=\theta_s(t+\alpha),\ \forall s\in\N, t\in\R^+.
\]
Together with the semi-group property of the first oscillator dynamics, in order to prove the existence of TW, it suffices to find a forcing signal $f$ and a phase shift $\alpha$ such that $\theta_1^{(f)}(t)=f(t+\alpha)$ for all $t\in\R^+$. Using the translation operator $R^{-\alpha}f(t)=f(t-\alpha)$, this is equivalent to solving the following delay-differential equation
\[
\frac{df}{dt}=1+\epsilon\Delta(f)\delta(R^{-\alpha}f),
\]
for the pair $(f,\alpha)$. The purpose of the section is precisely to solve this equation. To that goal, it is useful to begin by identifying all possible cases.

\begin{figure}[h]
\centerline{\includegraphics[scale=0.6]{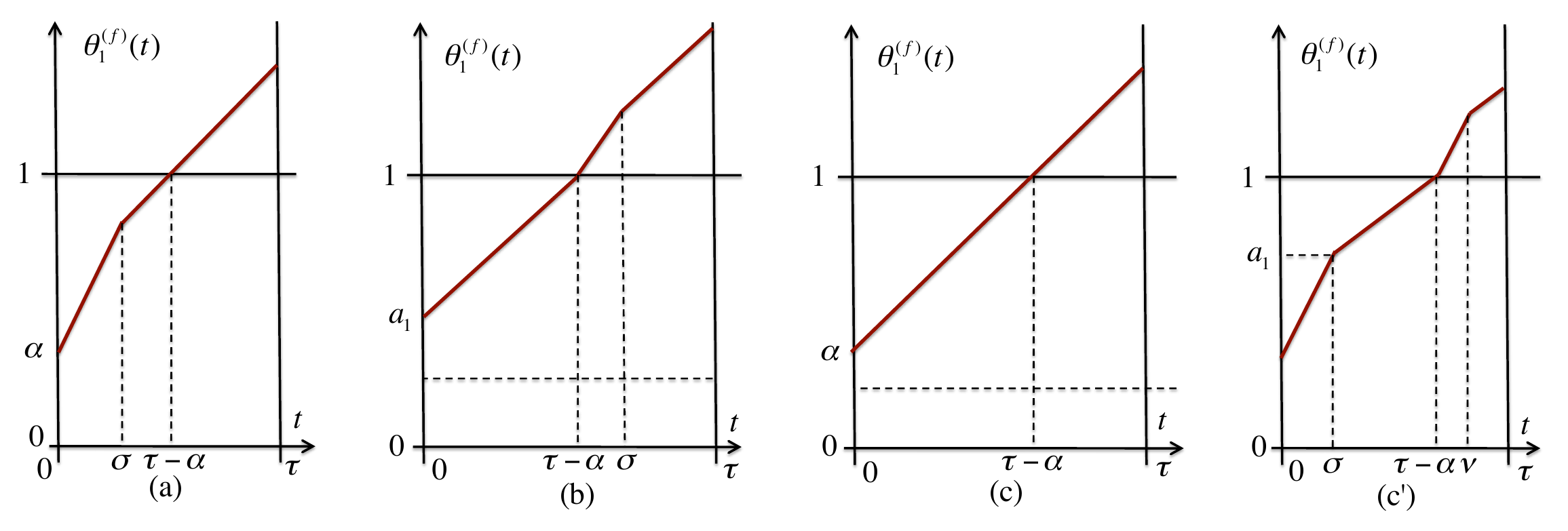}}
\caption{Examples of the function $t\mapsto \theta_1^{(f)}(t)$ for $t\in [0,\tau]$ in each TW case of Lemma \ref{SPECIFY}.}
\label{EXAMPL}
\end{figure}

Since we assume $f(0)=0$, the forcing signal/TW shape $f$ can be entirely characterized by the partition of $[0,\tau]$ into intervals where the slope is constant. Without loss of generality, we can also assume that $\alpha\in(0,\tau)$.\footnote{Of note, $f(0)=0$ implies $f(\tau)=1$ and we must have $f(\alpha)>0$.}
Under these assumptions, there can only be four cases depending on the initial location $f(\alpha)$ of $\theta_1$ (above or below $a_1$) and the location $f(\tau-\alpha)$ of $\theta_0$ (above or below $a_0$) when $\theta_1$ reaches 1. By examining each case, one easily constructs the desired partition.
\begin{Lem}
A TW shape $f$ must fall into one of the following cases (see Figure \ref{EXAMPL}):
\begin{itemize}
\item[{\rm (a)}] $f(\alpha)<a_1$ and $f(\tau-\alpha)\geq a_0$. Then there exists $\sigma\in (0,\tau)$ such that we have for the TW coordinate at site $s=1$
\[
\frac{d\theta_1^{(f)}}{dt}=\left\{\begin{array}{ccl}
1+\epsilon&\text{if}& 0<t<\sigma\\
1&\text{if}&\sigma<t<\tau
\end{array}\right.
\]
\item[{\rm (b)}] $f(\alpha)\geq a_1$ and $f(\tau-\alpha)< a_0$. Then there exists $\sigma\in (\tau-\alpha,\tau)$ such that we have
\[
\frac{d\theta_1^{(f)}}{dt}=\left\{\begin{array}{ccl}
1+\epsilon&\text{if}&\tau-\alpha<t<\sigma\\
1&\text{if}&0< t<\tau-\alpha\ \text{or if}\ \sigma<t\leq \tau
\end{array}\right.
\]
\item[{\rm (c)}] $f(\alpha)\geq a_1$ and $f(\tau-\alpha)\geq a_0$. Then $\frac{d\theta_1^{(f)}}{dt}=1$ for all $t\in [0,\tau]$.
\item[{\rm (c')}] $f(\alpha)<a_1$ and $f(\tau-\alpha)< a_0$. Then there exist $\sigma\in (0,\tau-\alpha)$ and $\nu\in (\tau-\alpha,\tau)$ such that we have
\[
\frac{d\theta_1^{(f)}}{dt}=\left\{\begin{array}{ccl}
1+\epsilon&\text{if}& 0\leq t<\sigma\quad \text{or if}\quad \tau-\alpha<t<\nu\\
1&\text{if}&\sigma<t<\tau-\alpha\quad \text{or if}\quad \nu<t<\tau
\end{array}\right.
\]
\end{itemize}
\label{SPECIFY}
\end{Lem}
{\sl Proof.} The four cases are clearly mutually exclusive. That they fully characterize the shape over the period (and the explicit expressions as claimed) is a consequence of the following observations.
\begin{itemize}
\item[(a)] The assumption $f(\alpha)<a_1$ implies that $\theta_1$ immediately increases with speed $1+\epsilon$ when $t$ crosses 0, and until either it reaches $a_1$ or $\theta_0$ reaches $a_0$. Then $\theta_1$ must increase with speed 1 until (at least) time $\tau-\alpha$ when it reaches 1. The inequality $f(\tau-\alpha)\geq a_0$ implies that $\theta_0$ is above $a_0$ for all $t>\tau-\alpha$; hence $\theta_1$ continues to grow at rate 1 until $\theta_1$ reaches 1. 
\item[(c')] Assuming a first phase as before, if otherwise $f(\tau-\alpha)<a_0$, then $\theta_1$ increases again with speed $1+\epsilon$ after it has crossed 1, and until either $\theta_0$ reaches $a_0$ or $\theta_1$ reaches $1+a_1$. The latter case is impossible because it would imply that $\theta_1$ increases by more than 1 over the fundamental time interval. This uniquely determines case (c') and we have $f(\sigma-\alpha)=a_1$ and $f(\nu)=a_0$. 
\item[(c)] If $f(\alpha)\geq a_1$, then $\theta_1$ cannot have speed 1 before it reaches 1. Since we assume that $\theta_0$ is already larger than $a_0$ when this happens, it follows that $\theta_1$ can never have speed $1+\epsilon$.
\item[(b)] Assuming a first phase as in (c), if $f(\tau-\alpha)<a_0$, then $\theta_1$ accelerates after time $\tau-\alpha$ and until either $\theta_0$ reaches $a_0$ or $\theta_1$ reaches $1+a_1$. After that, $\theta_1$ has speed 1 until either $\theta_0$ reaches 1 or $\theta_1$ reaches 2. Again the latter case is impossible because total increase over one period is at most 1.\hfill $\Box$
\end{itemize}

For our purpose here, it is enough to focus on case (a). (Indeed, Lemma \ref{STABILITYLEM} below shows that these are the only possibly asymptotically stable waves.) In this case, Lemma \ref{SPECIFY} indicates that the TW shape is completely determined by the numbers $\alpha,\sigma$ and $\tau$. Our next statement claims that $\sigma$ and $\tau$ are actually entirely determined by the phase shift $\alpha$, and therefore, so is the TW shape. 
\begin{Lem}
For every choice of parameters $\epsilon,a_0,a_1$ and every phase shift $\alpha\in\R^+$, there exists at most one TW solution in case (a).  
\label{PROUNIQCAS}
\end{Lem}
One can actually prove that a similar statement holds in cases (b), (c) and (c'). More importantly, this statement paves the way to uniqueness as stated in Theorem \ref{MAINRES1}.

\noindent
{\sl Proof of the Lemma.} Together with continuity, Lemma \ref{SPECIFY} implies that the solution $\theta_1^{(f)}$, on the fundamental interval $[0,\tau]$, is given as follows\footnote{That $\theta_1^{(f)}(0)=f(\alpha)=\alpha$ is a consequence of the fact that $f(\tau-\alpha)\geq a_0$ - see figures in cases (a) and (c).} 
\[
\theta_1^{(f)}(t)=\left\{\begin{array}{ccl}
\alpha+(1+\epsilon)t&\text{if}&0\leq t\leq \sigma\\
\alpha+\eps \sigma+t&\text{if}&\sigma\leq t\leq\tau
\end{array}\right.
\]
By translation, this determines the shape $f(t)=\theta_1^{(f)}(t-\alpha)$ over the interval $[\alpha,\tau+\alpha]$, {\sl i.e.}\
\[
f(t)=\left\{\begin{array}{ccl}
\alpha+(1+\epsilon)(t-\alpha)&\text{if}&\alpha\leq t\leq \sigma+\alpha\\
\alpha+\eps \sigma-\alpha+t&\text{if}&\sigma+\alpha\leq t\leq\tau+\alpha
\end{array}\right.
\]  
We must have $\sigma<\tau-\alpha$ ({\sl i.e.}\ $\sigma+\alpha< \tau$) because $f$ has to have speed $1$ when it reaches 1 (see Figure \ref{EXAMPL}). Hence $f(t)$ must growth with speed 1 for $t\in (\tau,\tau+\alpha)$ and hence for $t\in (0,\alpha)$ by periodicity. 
Consequently, on the interval $[0,\tau]$, the TW shape writes $f(t)=\Theta_{\alpha,\sigma}(t)$ where the 2-parameter family $\{\Theta_{\alpha,s}\}_{\alpha,s\in\R^+}$ is defined by 
\begin{equation}
\Theta_{\alpha,s}(t)=\left\{\begin{array}{ccl}
t&\text{if}&t\leq \alpha\\
\alpha+(1+\eps)(t-\alpha)&\text{if}&\alpha\leq t\leq s+\alpha\\
\eps s+t&\text{if}&s+\alpha\leq t
\end{array}\right.\ \forall t\in\R^+.
\label{FIRSTFAMILY}
\end{equation}
Notice that $\Theta_{\alpha,s}(0)=0$ for every $\alpha,s\in\R^+$. Moreover, we must have $\Theta_{\alpha,\sigma}(\tau)=1$ which yields $\tau=1-\epsilon \sigma$. Therefore, in order to prove the lemma, it suffices to prove uniqueness of $\sigma$ given a phase shift $\alpha$. This is the purpose of the next statement. 
\begin{Claim}
For every choice of parameters $\epsilon,a_0,a_1$ and every phase shift $\alpha\in\R^+$, the equation $\Theta_{\alpha,t_0}(t_0)=a_0$ (resp.\ $\Theta_{\alpha,t_1}(t_1+\alpha)=a_1$) has a unique positive solution denoted $t_0$ (resp.\ $t_1$). Moreover, we have $\sigma=\min\{t_0,t_1\}$.
\end{Claim}
Notice that the quantity $\sigma$ in this statement satisfies the inequalities $0\leq \sigma\leq 1-\epsilon\sigma=\tau$.
 
\noindent
{\sl Proof of the Claim.} Assuming $t_0\geq 0$, the equation $\Theta_{\alpha,t_0}(t_0)=a_0$ has unique solution 
\[
\left\{\begin{array}{l}
t_0=a_0\ \text{if}\ a_0\leq \alpha\\
t_0=\frac{a_0+\alpha\epsilon}{1+\epsilon}\ \text{if}\ a_0>\alpha
\end{array}\right.\quad \text{that is}\quad t_0=\min\left\{a_0,\frac{a_0+\alpha\epsilon}{1+\epsilon}\right\}\geq 0.
\]
Assuming $t_1\geq 0$, the equation $\Theta_{\alpha,t_1}(t_1+\alpha)=a_1$ gives $t_1=\frac{a_1-\alpha}{1+\epsilon}$.

Now, $\sigma$ is the first time in $[0,\tau]$ when $\theta_1$ adopts speed $1$. As argued in the proof of Lemma \ref{SPECIFY}, this time corresponds to the smallest of times when $\theta_1$ reaches $a_1$ or $\theta_0$ reaches $a_0$; {\sl viz.}\ $\sigma=\min\{t_0,t_1\}$ and the Claim is proved.  \hfill $\Box$

To conclude about existence of waves, it remains to investigate the conditions on $\alpha$ such that a TW shape $\Theta_{\alpha,\sigma}$ satisfies the conditions of Lemma \ref{PROUNIQCAS}. The result is given in the next statement.
\begin{Lem}
There exists a TW solution in case (a) iff 
\[
0\leq\alpha<a_1\quad\text{and}\quad
0\leq\alpha\leq\max\left\{(1+\epsilon)(1-a_0)-\epsilon a_1,\frac{1-a_0+\epsilon (1-a_1)}{1+\epsilon}\right\}.
\]
\label{EXIST}
\end{Lem}
The constraints here define a non-empty interval of $\alpha$ for every possible choice of parameters $\epsilon,a_0,a_1$. By Lemma \ref{PROUNIQCAS}, let $I_{\epsilon,a_0,a_1}$ be the corresponding interval of forcing periods $\tau$. We have proved the following statement. 
\begin{Cor}
Equation \eqref{DEFDYNAM} has a unique traveling wave solution in case (a), for every forcing period in $I_{\epsilon,a_0,a_1}$.
\label{EXISTW}
\end{Cor}
{\sl Proof of Lemma \ref{EXIST}.} We need to check the conditions $0<f(\alpha)<a_1$ and $f(\tau-\alpha)\geq a_0$ of Lemma \ref{SPECIFY} for a shape $\Theta_{\alpha,\sigma}$ as in the proof of Lemma \ref{PROUNIQCAS}. 

First, we have $f(\alpha)=\Theta_{\alpha,\sigma}(\alpha)=\alpha$. So the condition $0<f(\alpha)<a_1$ is equivalent to $0<\alpha< a_1$.

In order to check the inequality $f(\tau-\alpha)\geq a_0$, we need to take into account the quantities $t_0$ and $t_1$ defined in the proof of Lemma \ref{PROUNIQCAS}. Notice that the condition $t_1<t_0$ is equivalent to 
\begin{equation}
\alpha>\alpha_0:=\max\left\{a_1-(1+\epsilon)a_0,\frac{a_1-a_0}{1+\epsilon}\right\}.
\label{ALPHA0}
\end{equation}
We consider the cases $\alpha\leq \alpha_0$ and $\alpha>\alpha_0$ separately. 

In the first case, the inequality $f(\tau-\alpha)\geq a_0$ is equivalent to $\tau-\alpha\geq t_0=\sigma$. However, we know from the proof of Lemma \ref{PROUNIQCAS} that we must have $\tau-\alpha>\sigma$; hence there is nothing to prove.

In the second case $\alpha>\alpha_0$, we first observe that $\Theta_{\alpha,s}(t_0)=a_0$ for all $s\in\R^+$ and in particular for $s=t_1$. Therefore, and using also $\tau=1-\epsilon\sigma=1-\epsilon t_1$ from the proof of Lemma \ref{PROUNIQCAS}, the inequality $f(\tau-\alpha)\geq a_0$ is equivalent to $1-\epsilon t_1-\alpha\geq t_0$, that is to say 
\[
\alpha\leq\alpha_1:=\max\left\{(1+\epsilon)(1-a_0)-\epsilon a_1,1-\frac{a_0+\epsilon a_1}{1+\epsilon}\right\}.
\]
Altogether, we conclude that the inequality $f(\tau-\alpha)\geq a_0$ is equivalent to $\alpha\leq \max\{\alpha_0,\alpha_1\}$. Elementary algebra, together with the assumption $a_1<1$, show that $\alpha_0\leq \alpha_1$; hence the desired condition follows. \hfill $\Box$ 

\section{Stability analysis}\label{S-STAB}
This section reports the stability analysis of TW and aims to establish the first statement in Theorem \ref{MAINRES2}. We first focus on the stability analysis, first local and then global, of fixed points of the {\bf stroboscopic map} 
$F_f:\T\to\T$ that updates and translates back the first coordinate $\theta_1$ after a forcing period, namely 
\begin{equation*}
F_f(\vartheta_1):=\theta_1^{(f)}(\vartheta_1,\tau)-1.
\end{equation*}
By periodicity of $f$ and of the 'vector field' in equation \eqref{DEFDYNAM}, we have 
\[
F_f^n(\vartheta_1)=\theta_1^{(f)}(\vartheta_1,n\tau)-n,\ \forall n\in\N,
\]
hence, the orbits of $F_f$ indeed capture the asymptotic behavior of $\theta_1$. 
Equation \eqref{DEFDYNAM} implies that $F_f$ is a lift of an endomorphism of the circle that preserves orientation.\footnote{Indeed, we obviously have $F_f(x+1)=F_f(x)+1$ for all $x$, from equation \eqref{DEFDYNAM}. Moreover, the continuous dependence of $\theta_1^{(f)}(\vartheta_1,\tau)$ on $\vartheta_1$ implies that $F_f$ must be continuous. As for monotonicity, by contradiction, if we had $\vartheta_1< \xi_1$ and $F_f(\vartheta_1)>F_f(\xi_1)$, then the corresponding trajectories $t\mapsto \theta_1^{(f)}(\vartheta_1,t)$ and $t\mapsto \theta_1^{(f)}(\xi_1,t)$  would have crossed for some $t\in (0,\tau)$. This is impossible by uniqueness of the vector field acting on this coordinate.} One can show that $F_f$ is actually strictly increasing; hence it is a homeomorphism from $\T$ into itself.

The theory of circle maps (see e.g.\ \cite{KH95}) states the existence of a rotation number for every orbit, namely the following limit exists
\[
\lim_{n\to+\infty}\frac{F_f^n(\vartheta_1)}{n},\ \forall \vartheta_1\in\T
\]
and does not depend on $\vartheta_1$. In principle, this number can take any value in $\R$. However, a TW solution exists iff the corresponding stroboscopic map has a fixed point, that is to say iff
\[
F_f(f(\alpha))=f(\alpha).
\]
In particular, the stroboscopic map associated with a TW must have vanishing rotation number. 

\subsection{Local stability of stroboscopic map fixed points}
In this section, we study the local stability of the fixed point $f(\alpha)$ of the stroboscopic map $F_f$ associated with a TW shape. 
A first statement respectively identifies the stable, unstable and neutral cases according to the decomposition in Lemma \ref{SPECIFY}.
\begin{Lem}
Following the decomposition in Lemma \ref{SPECIFY}, we have
\begin{itemize}
\item[$\bullet$] the fixed point $f(\alpha)$ is locally asymptotically stable if in case (a) with $f(\sigma)<a_0$ when $\theta_1^{(f)}(\sigma)=a_1$. 
\item[$\bullet$] It is unstable if in case (b) with $\theta_1^{(f)}(\sigma)<1+a_1$ when $f(\sigma)=a_0$, 
\item[$\bullet$] and is neutral in any other case.
\end{itemize} 
\label{STABILITYLEM}
\end{Lem}
{\sl Proof.} We want to evaluate the behavior of the difference $F_f^n(f(\alpha))-F_f^n(\xi_1)$, where $\xi_1\in\T$ is a small perturbation of the initial condition $\vartheta_1=f(\alpha)$. 

{\em Stable case.} By continuous dependence of $\theta_1^{(f)}(\vartheta_1,t)$ on $\vartheta_1$, the time $\sigma'$ when $\theta_1^{(f)}(\xi_1,\sigma')=a_1$ can be made arbitrarily close to $\sigma$ by choosing $\xi_1$ sufficiently close to $f(\alpha)$. Therefore, there exists $\delta>0$ such that for every $\xi_1\in (f(\alpha)-\delta,f(\alpha))$, there exists $\sigma'>\sigma$ such that $f(\sigma')<a_0$ when $\theta_1^{(f)}(\xi_1,\sigma')=a_1$. Accordingly, the dynamics of the two trajectories in the interval $[0,\tau]$ can be summarized as follows:
\begin{itemize}
\item[$\bullet$] both $\theta_1^{(f)}(\vartheta_1,t)$ and $\theta_1^{(f)}(\xi_1,t)$ have speed $1+\epsilon$ for $t\in (0,\sigma)$.
\item[$\bullet$] $\theta_1^{(f)}(\vartheta_1,t)$ has speed 1 and $\theta_1^{(f)}(\xi_1,t)$ has speed $1+\epsilon$ for $t\in (\sigma,\sigma')$.
\item[$\bullet$] both $\theta_1^{(f)}(\vartheta_1,t)$ and $\theta_1^{(f)}(\xi_1,t)$ have speed $1$ for $t\in (\sigma',\tau)$.
\end{itemize}
It results that $\theta_1^{(f)}(\xi_1,\cdot)$ has speed $1+\epsilon$ on a longer time interval and thus 
\[
f(\alpha)-F_f(\xi_1)=F_f(f(\alpha))-F_f(\xi_1)<f(\alpha)-\xi_1.
\] 
We also have $F_f(f(\alpha))-F_f(\xi_1)> 0$ by strict monotonicity; hence 
\[
F_f(\xi_1)\in (f(\alpha)-\delta,f(\alpha)),
\]
and the argument can be repeated with $F_f(\xi_1)$ to obtain
\[
0< f(\alpha)-F_f^2(\xi_1)<f(\alpha)-F_f(\xi_1)<\delta.
\]
By induction, it follows that the sequence $\{f(\alpha)-F_f^n(\xi_1)\}_{n\in\N}$ is decreasing and non-negative; hence it converges. A standard contradiction argument based on the contraction and on the continuity of $F_f$ concludes that the limit must be 0. This proves local asymptotic stability  with respect to negative initial perturbations. A similar argument applies to positive perturbations. 

{\em Unstable case.} Similarly to as before, let $\delta>0$ be sufficiently small so that for every $\xi_1\in (f(\alpha)-\delta,f(\alpha))$, we have $\theta_1^{(f)}(\xi_1,\sigma')<1+a_1$. Comparing again the two trajectories, we have:
\begin{itemize}
\item[$\bullet$] both $\theta_1^{(f)}(\vartheta_1,t)$ and $\theta_1^{(f)}(\xi_1,t)$ have speed $1$ for $t\in (0,\tau-\alpha)$. (NB: $\tau-\alpha=1-f(\alpha)$).
\item[$\bullet$] $\theta_1^{(f)}(\vartheta_1,t)$ has speed $1+\epsilon$ and $\theta_1^{(f)}(\xi_1,t)$ has speed $1$ for $t\in (\tau-\alpha,1-\xi_1)$.
\item[$\bullet$] both $\theta_1^{(f)}(\vartheta_1,t)$ and $\theta_1^{(f)}(\xi_1,t)$ have speed $1+\epsilon$ for $t\in (1-\xi_1,\sigma)$.
\item[$\bullet$] both $\theta_1^{(f)}(\vartheta_1,t)$ and $\theta_1^{(f)}(\xi_1,t)$ have speed $1$ for $t\in (\sigma,\tau)$.
\end{itemize}
In this case, $\theta_1^{(f)}(\xi_1,\cdot)$ has speed $1+\epsilon$ on a shorter interval and thus $F_f(f(\alpha))-F_f(\xi_1)>f(\alpha)- \xi_1$ for every $\xi_1\in (f(\alpha)-\delta,f(\alpha))$, {\sl viz.}\ the TW is unstable with respect to negative perturbations. A similar argument applies to positive perturbations.

{\em Neutral case.} The analysis is similar. One shows that the total duration when the perturbed trajectory has speed $1+\epsilon$ is identical to that of the TW; hence $F_f(f(\alpha))-F_f(\xi_1)=f(\alpha)-\xi_1$. \hfill $\Box$

In the proof of Lemma \ref{EXIST} above, we have identified the condition $\alpha>\alpha_0$ (see equation \eqref{ALPHA0}) as sufficient to ensure being in case (a) with $f(\sigma)<a_0$ when $\theta_1^{(f)}(\sigma)=a_1$. By cross-checking this constraint with the one in the statement of that Lemma (and using also Lemma \ref{PROUNIQCAS}, {\sl i.e.}\ that the TW is entirely determined by the parameters $\epsilon,a_0,a_1$ and its phase shift $\alpha$), we get the following conclusion.
\begin{Cor}
For every choice of parameters $\epsilon,a_0,a_1$, there exists an interval $I'_{\epsilon,a_0,a_1}\subset I_{\epsilon,a_0,a_1}$ such that for every $\tau\in I'_{\epsilon,a_0,a_1}$, there exists a $\tau$-periodic TW shape $f$ and a phase shift $\alpha\in (0,\tau)$ such that $f(\alpha)$ is a locally asymptotically stable fixed point of $F_f$.
\label{GLOBEXISTS}
\end{Cor}
{\sl Proof.} In order to be able to choose $\alpha$ that simultaneously satisfies $\alpha>\alpha_0$ and the conditions of Lemma \ref{EXIST}, it suffices to prove that the inequality 
\[
\max\left\{0,a_1-(1+\epsilon)a_0,\frac{a_1-a_0}{1+\epsilon}\right\}<\min\left\{a_1,\max\left\{(1+\epsilon)(1-a_0)-\epsilon a_1,\frac{1-a_0+\epsilon (1-a_1)}{1+\epsilon}\right\}\right\},
\]
holds for all parameter values. To that goal, we consider separately different parameter regimes. 

\noindent
If $a_1\leq a_0$, then the left hand side of the inequality is equal 0, while the right hand side always remains non-negative; hence the inequality holds.

\noindent
In order to investigate the case $a_1>a_0$, we notice that direct calculations imply the following conclusions
\begin{itemize}
\item[$\bullet$] the inequality $\frac{a_1-a_0}{1+\epsilon}\leq a_1-(1+\epsilon)a_0$ is equivalent to $(2+\epsilon)a_0\leq a_1$,
\item[$\bullet$] the condition $a_1\leq \max\left\{(1+\epsilon)(1-a_0)-\epsilon a_1,\frac{1-a_0+\epsilon (1-a_1)}{1+\epsilon}\right\}$ simplifies to $a_1\leq 1-\min\left\{a_0,\frac{a_0+\epsilon a_1}{1+\epsilon}\right\}$ and we obviously have $\min\left\{a_0,\frac{a_0+\epsilon a_1}{1+\epsilon}\right\}=a_0$ when $a_1>a_0$.
\end{itemize}
Furthermore, we clearly have $\max \left\{ 0 , \frac{a_1-a_0}{1+\eps} , a_1 - (1+ \eps )a_0 \right\}\leq a_1$; hence the statement also holds in the case where $a_0<a_1\leq 1-a_0$.

\noindent
From now on, we can assume $a_1\geq \max\{a_0,1-a_0\}$. We have 
\[
\frac{1-a_0+\epsilon (1-a_1)}{1+\epsilon}\leq (1+\epsilon)(1-a_0)-\epsilon a_1\ \Longleftrightarrow
(2+\epsilon)a_0+\epsilon a_1\leq 1+\epsilon.
\]
Accordingly, one has to consider three cases
\begin{itemize}
\item[$\bullet$] if $(2+\epsilon)a_0\leq a_1$ then $(2+\epsilon)a_0+\epsilon a_1\leq 1+\epsilon$ and the statement holds provided that $a_1-(1+\epsilon)a_0\leq (1+\epsilon)(1-a_0)-\epsilon a_1$, which is true.
\item[$\bullet$] If $(2+\epsilon)a_0+\epsilon a_1> 1+\epsilon$ then $(2+\epsilon)a_0> a_1$ and the inequality to check is 
\[
\frac{a_1-a_0}{1+\eps}\leq \frac{1-a_0+\epsilon (1-a_1)}{1+\epsilon},
\]
which holds true.
\item[$\bullet$] Finally, if $(2+\epsilon)a_0> a_1$ and $(2+\epsilon)a_0+\epsilon a_1\leq 1+\epsilon$, then the inequality to verify is 
\[
\frac{a_1-a_0}{1+\eps}\leq (1+\epsilon)(1-a_0)-\epsilon a_1,
\]
which is equivalent to $a_1-a_0\leq (1+\epsilon)\left(1+\epsilon(1-a_1)-(1+\epsilon)a_0\right)$. However, the inequality $(2+\epsilon)a_0+\epsilon a_1\leq 1+\epsilon$ implies $(1+\epsilon)\left(1+\epsilon(1-a_1)-(1+\epsilon)a_0\right)\geq (1+\epsilon)a_0$ and we have $(1+\epsilon)a_0\geq a_1-a_0$ because of $(2+\epsilon)a_0> a_1$. The proof of the Corollary is complete. \hfill $\Box$
\end{itemize}

\subsection{Global stability of fixed points}
Once local stability has been established, a careful computation of $F_f$ allows one to show that the fixed points are actually globally stable.
\begin{Pro}
When $\tau\in I'_{\epsilon,a_0,a_1}$, the fixed point $f(\alpha)$ is globally stable. That is to say, there exists a unique unstable fixed point $\vartheta_\text{unst}\in \T$ such that we have 
\[
\lim_{n\to+\infty}F_f^n(\vartheta_1)=f(\alpha)\ (\text{\rm mod}\ 1),\ \forall \vartheta_1\in\T\ :\ \vartheta_1\neq \vartheta_\text{unst}.
\]
\label{GLOBSTAB}
\end{Pro}
To be more accurate, the proof below actually shows that for every $\vartheta_1\neq \vartheta_\text{unst}$, there exists $m_{\vartheta_1}\in\Z$ such that
\[
\lim_{n\to+\infty}F_f^n(\vartheta_1)=f(\alpha)+m_{\vartheta_1}.
\]

{\sl Proof.} We are going to prove that, for the TW in case (a) with $\alpha$ simultaneously satisfying $\alpha>\alpha_0$ and the conditions of Lemma \ref{EXIST}, the restriction of $F_f$ to the interval $[0,1)$ consists of four affine pieces (each piece being defined on an interval): two pieces are rigid rotations and they are interspersed by one contracting and one expanding piece. 

Since $F_f$ is a lift of an endomorphism of the circle that preserves the orientation, and since it has a locally stable fixed point $f(\alpha)$ (Corollary \ref{GLOBEXISTS}), the graph of the contracting piece must intersect the main diagonal of $\R^2$. As a consequence, the graphs of the following and preceding neutral pieces cannot intersect this line. By continuity and periodicity, the graph of the remaining expanding piece must intersect this line as well. Let $\vartheta_\text{unst}\in [0,1)$ be this unique unstable fixed point. The Proposition immediately follows.

In order to prove the decomposition into the four desired pieces, we are going to consider various cases. To that goal, recall first that $f(0)=0$ and $f(t_0)=a_0$, {\sl i.e.}\ $\theta_1^{(f)}$ can only have speed $1+\epsilon$ for $t< t_0$ (within the interval $[0,\tau)$). Consider the trajectory of the coordinate $\theta_1$ with initial value $\vartheta_1=0$. Since we have $F_f(0)<F_f(\alpha)=\alpha<1$, the largest integer $\theta_1$ reaches before time $t_0$ is at most 1, {\sl viz.}\
\[
n=\max\left\{k\in\Z^+:k(\frac{a_1}{1+\epsilon}+1-a_1)\leq t_0\right\}\in\{0,1\},
\]
We consider separately two cases:
\begin{itemize}
\item[(a)] either the last rapid phase with speed $1+\epsilon$ (when $\theta_1\geq n$) stops when $t=t_0$. This occurs iff $t_0< n(\frac{a_1}{1+\epsilon}+1-a_1)+\frac{a_1}{1+\epsilon}$
\item[(b)] or its stops when $\theta_1$ reaches $n+a_1$.\footnote{possibly at $t=t_0$, {\sl i.e.}\ we may have $\theta_1(t_0)=n+a_1$.} This occurs when $ n(\frac{a_1}{1+\epsilon}+1-a_1)+\frac{a_1}{1+\epsilon}\leq t_0<(n+1)(\frac{a_1}{1+\epsilon}+1-a_1)$
\end{itemize}

\begin{figure}[h]
\centerline{\includegraphics[scale=0.6]{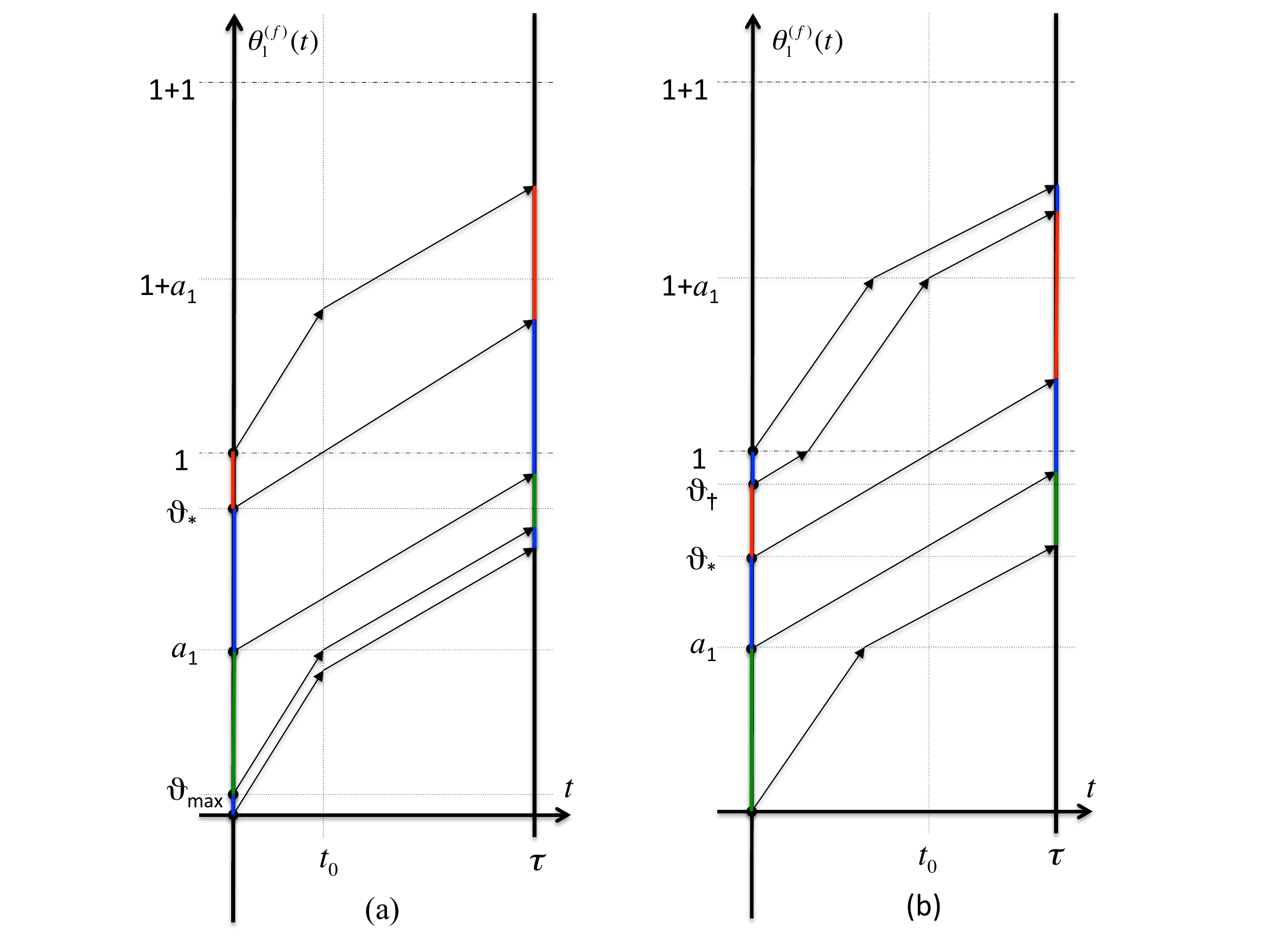}}
\caption{Illustrative examples of the four-interval decomposition of the stroboscopic map $F_f$, as discussed in the two major cases of Proposition \ref{GLOBSTAB} (for $n=0$ and $a_1<\vartheta_\ast$). The dynamics of critical initial conditions are shown in light solid black. The blue-colored intervals denote the subset of $\T$ on which $F_f$ is a translation, whereas green and red denote contraction and expansion, respectively.}
\label{F-FOUR_INTERVALS}
\end{figure}

See Figure \ref{F-FOUR_INTERVALS} for an illustration of the action of $F_f$ according to these two cases, when $n=0$.

Throughout the proof, we shall  
frequently make use of the time $t_{a,\vartheta_1}$ when the coordinate $\theta_1$ reaches the value $a$, {\sl i.e.}\ $\theta_1^{(f)}(\vartheta_1,t_{a,\vartheta_1})=a$. Here $0\leq \vartheta_1\leq a$ are arbitrary.

{\sl Case (a).} By continuous and monotonic dependence on initial conditions, every trajectory starting initially with $\vartheta_1$ sufficiently close to 0 (and $>0$), will not only experience the same number $n+1$ of rapid phases with speed $1+\epsilon$, but the last rapid phase (the unique rapid phase if $n=0$) will also stop at $t=t_0$. 

Assume $n=1$. As time evolves between 0 and $t_0$, speed changes for such trajectories occur when the level lines $a_1$ and $1$ are reached ({\sl i.e.}\ when $t=t_{a_1,\vartheta_1}$ and $t=t_{1,\vartheta_1}$). Consequently, the delays between the corresponding instants for two distinct trajectories remain constants {\sl i.e.}\ we have 
\[
t_{a_1,0}-t_{a_1,\vartheta_1}=t_{1,0}-t_{1,\vartheta_1}
\]
This equality implies that the cumulated lengths of the rapid phases satisfy
\[
t_{a_1,0}+t_0-t_{1,0}=t_{a_1,\vartheta_1}+t_0-t_{1,\vartheta_1}
\]
{\sl i.e.}\ they are independent of $\vartheta_1$. We conclude that $F_f(\vartheta_1)=F_f(0)+\vartheta_1$; {\sl viz.}\ the map $F_f$ is a rigid rotation in some right neighborhood of 0. 

If $n=0$, the same conclusion immediately follows from the fact that lengths of the rapid phases are simply given by $t_0$.

Moreover, the largest initial condition $\vartheta_\text{max}$ for which this property holds is defined by $\theta_1(\vartheta_\text{max},t_0)=n+a_1$ and we have $\vartheta_\text{max}\leq a_1$. Therefore, $F_f$ is a rigid rotation on $(0,\vartheta_\text{max})$. To continue, we separate case (a) into 2 subcases:
\begin{itemize}
\item[(a1)] either $\theta_1^{(f)}(a_1,t_0)\leq n+1$.
\item[(a2)] or $\theta_1^{(f)}(a_1,t_0)> n+1$
\end{itemize}
Assume first that case (a1) holds. In the case $n=1$, using similar considerations as above, for trajectories now starting in $[\vartheta_\text{max},a_1]$, we obtain
\[
t_{a_1,\vartheta_\text{max}}-t_{a_1,a_1}=t_{1,\vartheta_\text{max}}-t_{1,a_1}=t_{1+a_1,\vartheta_\text{max}}-t_{1+a_1,a_1}
\]
and $t_{a_1,a_1}=0$ and $t_0=t_{1+a_1,\vartheta_\text{max}}$ (and monotonicity together with $\vartheta_\text{max}\leq a_1$ implies the existence of $t_{1+a_1,a_1}\leq t_0$). For the cumulated lengths of rapid phases, the last equality implies  
\[
t_{1+a_1,\vartheta_\text{max}}-t_{1,\vartheta_\text{max}}=t_{1+a_1,a_1}-t_{1,a_1}
\]
and using that for the initial rapid phase (which is the only rapid phase if $n=0$), we have 
\[
\theta_1^{(f)}(\vartheta_\text{max},t_{a_1,\vartheta_\text{max}})=(a_1=)\ \vartheta_\text{max}+t_{a_1,\vartheta_\text{max}}(1+\epsilon)\quad\text{and}\quad  \theta_1^{(f)}(a_1,t_{a_1,\vartheta_\text{max}})=a_1+t_{a_1,\vartheta_\text{max}},
\]
we obtain that $F_f(a_1)=F_f(\vartheta_\text{max})+a_1-\vartheta_\text{max}-\epsilon t_{a_1,\vartheta_\text{max}}$ from where it follows using that $F_f$ is piecewise affine, that the restriction of $F_f$ to the interval $(\vartheta_\text{max},a_1)$ must be a contraction. 

Let $\vartheta_\ast$ be such that $\theta_1^{(f)}(\vartheta_\ast,t_0)=n+1$. The dichotomy (a1) {\sl vs.} (a2) is equivalent to $a_1\leq \vartheta_\ast$ {\sl vs.}\ $a_1>\vartheta_\ast$. In the latter case, the same conclusion as before applies to the interval $(\vartheta_\text{max},\vartheta_\ast)$.   

Now, for $\vartheta_1\in (a_1,\vartheta_\ast)$ when in case (a1) and $n=1$, the equality between change speed instants writes
\[
t_{1,\vartheta_\ast}-t_{1,a_1}=t_{1+a_1,\vartheta_\ast}-t_{1+a_1,a_1}
\]
and similar calculations to those for $\vartheta_1\in (0,\vartheta_\text{max})$ show that $F_f$ is also a rigid rotation on $(a_1,\vartheta_\ast)$. 

If $n=0$, that $F_f$ is a rigid rotation on $(a_1,\vartheta_\ast)$ immediately follows from the fact that there is no rapid phase at all. In case (a2), the same conclusion applies to $(\vartheta_\ast,a_1)$.

Finally, for the interval $(\vartheta_\ast,1)$ in case (a1), one shows that the cumulated rapid phases for the trajectories issued from $\vartheta_\ast$ and $1$ are identical up to $t=t_{n+1,1}$. Moreover, on the interval $(t_{n+1,1},t_0)$, the function $\theta_1^{(f)}(\vartheta_\ast,\cdot)$ has speed 1 and $\theta_1^{(f)}(1,\cdot)$ has speed $1+\epsilon$ (because $\theta_1^{(f)}(t_0,1)\in n+1+(0,a_1)$ by degree 1).\footnote{In particular, this shows that $(\vartheta_\ast,1)$ is the fourth and last interval to consider in this analysis. There are no more acceleration pattern and $F_f$ indeed decomposes over exactly four intervals.}  It follows that
\[
F_f(1)=F_f(\vartheta_\ast)+1-\vartheta_\ast+\epsilon (t_0-t_{n+1,1})
\]
which implies $F_f$ must be an expansion on this interval. In case (a2), the same conclusion easily follows for the interval $(a_1,1)$.

{\sl Case (b).} The arguments are similar. For $\vartheta_1>0$ in the neighborhood of 0, the length of the last rapid phase is now independent of $\vartheta_1$ ($n=1$) and the length of the first rapid phase decreases when $\vartheta_1$ increases. Hence $F_f$ is contraction on this first interval. Its upper boundary is given by $\min\{a_1,\vartheta_\ast\}$, where as before $\vartheta_\ast$ is defined by $\theta_1(\vartheta_\ast,t_0)=n+1$.  

For $\vartheta_1\in (a_1,\vartheta_\ast)$ (or $\vartheta_1\in (\vartheta_\ast,a_1)$ depending on the case), the cumulated duration of the accelerated phases does not depend on $\vartheta_1$ (in particular because the last rapid phase stops when the coordinate reaches $n+a_1$). 

In the case $a_1<\vartheta_\ast$, for $\vartheta_1>\vartheta_\ast$ the trajectory has an extra final rapid phase when compared to $\vartheta_\ast$ and this holds for all $\vartheta_1\leq \vartheta_\dag$ where $\vartheta_\dag$ is defined by $\theta_1(\vartheta_\dag,t_0)=n+1+a_1$.\footnote{The existence of $\vartheta_\dag$ is granted from the fact that $\theta_1(1,t_0)> n+1+a_1$ which follows from $\theta_1(0,t_0)>n+a_1$.} Finally, for $\vartheta_1\in (\vartheta_\dag,1)$, $F_f$ is a rigid rotation. The case $\vartheta_\ast\leq a_1$ can be treated similarly. \hfill $\Box$

\subsection{Global stability of TW: proof of Theorem \ref{MAINRES15}}
In order to prove Theorem \ref{MAINRES15}, we first consider the stroboscopic dynamics at the second site and then extend the results down the chain. Thanks to the structure of equation \eqref{DEFDYNAM}, the second site coordinate after one forcing period can be expressed in terms of the time-$\tau$ stroboscopic map $F_{\theta_1^{(f)}(\vartheta_1,\cdot)}$ as follows
\[
\theta_2^{(f)}(\vartheta_1,\vartheta_2,\tau)=F_{\theta_1^{(f)}(\vartheta_1,\cdot)}(\vartheta_2)+1,
\]
By induction, we have (recall that $R^\tau$ represents the time translation by an amount $\tau$, {\sl i.e.}\ $R^\tau\theta (\cdot)=\theta (\cdot+\tau)$)
\[
\theta_2^{(f)}(\vartheta_1,\vartheta_2,n\tau)-n=F_{f,\vartheta_1}^{(n)}(\vartheta_2):=F_{R^{n\tau}\theta_1^{(f)}(\vartheta_1,\cdot)}\circ \cdots\circ F_{R^\tau\theta_1^{(f)}(\vartheta_1,\cdot)}\circ F_{\theta_1^{(f)}(\vartheta_1,\cdot)}(\vartheta_2),\ \forall n\in\N.
\]
The semi-group property of the flow implies that $f(2\alpha)$ is a globally stable fixed point of the map $F_{R^\alpha f}$.
Using similar arguments as in the stability proof in \cite{LM14}, we use this property here to show that this stable point attracts the iterates $F_{f,\vartheta_1}^{(n)}(\vartheta_2)$.
\begin{Pro}
Let $\tau\in I'_{\epsilon,a_0,a_1}$, $\theta$ be the associated case (a) TW shape and $\vartheta_\text{unst}$ be given by Proposition \ref{GLOBSTAB}. For every $\vartheta_1\neq\vartheta_\text{unst}$, there exists a unique $\vartheta_\text{unst,2}\in\T$ such that we have 
\[
\lim_{n\to+\infty}F_{f,\vartheta_1}^{(n)}(\vartheta_2)=f(2\alpha)\ (\text{\rm mod}\ 1),\ \forall \vartheta_2\in\T\ :\ \vartheta_2\neq \vartheta_\text{unst,2}.
\]
\label{STAB2SITE}
\end{Pro}
The same comment as the one after Proposition \ref{GLOBSTAB} applies here. Namely, the proof actually establishes that that for every $\vartheta_2\neq \vartheta_\text{unst,2}$, there exists $m_{\vartheta_2}\in\Z$ such that
\[
\lim_{n\to+\infty}F_{f,\vartheta_1}^{(n)}(\vartheta_2)=f(2\alpha)+m_{\vartheta_2}.
\]

{\sl Proof.} Let $\vartheta_1\neq \vartheta_\text{unst}$ be fixed. The proof relies on the following property of the stroboscopic map.
\begin{itemize}
\item[(P)] We have $\lim\limits_{n\to+\infty}\|F_{R^{n\tau}\theta_1^{(f)}(\vartheta_1,\cdot)}-F_{R^\alpha f}\|=0$ where the uniform norm $\|\cdot\|$ is extended to functions on $\T$.
\end{itemize}
In order to prove this fact, notice that Proposition \ref{GLOBSTAB} states that $R^{n\tau}\theta_1^{(f)}(\vartheta_1,0):=\theta_1^{(f)}(\vartheta_1,n\tau)$ converges to $R^\alpha f(0)\ (\text{mod}\ 1)$. When regarding these points as initial conditions in $\T$ for the one-dimensional system submitted to stimulus $f$, the continuous dependence of solutions on initial conditions implies 
\[
\lim_{n\to+\infty}\|R^{n\tau}\theta_1^{(f)}(\vartheta_1,\cdot)|_{[0,\tau]}\|=\|R^\alpha f|_{[0,\tau]}\|.
\]
The convergence $F_{R^{n\tau}\theta_1^{(f)}(\vartheta_1,\cdot)}(\vartheta_2)\to F_{R^\alpha f}(\vartheta_2)$ for every $\vartheta_2$ then follows from the continuous dependence of solutions on the forcing signal. Finally, that the convergence is uniform in $\T$ is a consequence of the continuity and monotonicity of the maps $F_{R^{n\tau}\theta_1^{(f)}(\vartheta_1,\cdot)}$ and $F_{R^\alpha f}$.

Now, the first step of the proof consists in showing that every sequence $\{F_{f,\vartheta_1}^{(n)}(\vartheta_2)\}_{n\in\N}$ either converges to a point $f(2\alpha)\ (\text{\rm mod}\ 1)$ or to one point in $f(t_\text{unst}+\alpha)\ (\text{\rm mod}\ 1)$ where $t_\text{unst}\in [0,\tau)$ is defined by 
\[
f(t_\text{unst})\ (\text{\rm mod}\ 1)=\vartheta_\text{unst}.
\]
To that goal, we are going to prove that if a sequence does not converge to $f(t_\text{unst}+\alpha)\ (\text{\rm mod}\ 1)$, {\sl i.e.}\ if there exists $\delta>0$ and a diverging subsequence $\{n_k\}_{k\in\N}$ so that the following estimate for distances in $\T$ holds
\begin{equation}
|F_{f,\vartheta_1}^{(n_k)}(\vartheta_2)-f(t_\text{unst}+\alpha)\ (\text{\rm mod}\ 1)|>\delta,\ \forall k\in\N,
\label{BOUNDEDAWAY}
\end{equation}
then we have
\[
\lim_{n\to+\infty}F_{f,\vartheta_1}^{(n)}(\vartheta_2)=f(2\alpha)\ (\text{\rm mod}\ 1).
\]

Together with periodicity, the inequality \eqref{BOUNDEDAWAY} implies the existence $\bar\delta\in (0, |f(t_\text{unst}+\alpha)-f(2\alpha)\ (\text{\rm mod}\ 1)|)$ such that 
\[
|F_{f,\vartheta_1}^{(n_k)}(\vartheta_2)-f(2\alpha)\ (\text{\rm mod}\ 1)|<\bar\delta,\ \forall k\in\N.
\]
The behavior of the map $F_{R^\alpha f}$ in the interval $f(2\alpha)+(-\bar\delta,\bar\delta)$ around the stable fixed point $f(2\alpha)$ implies the existence of $\eta\in (0,\bar\delta)$ such that 
\begin{equation}
|F_{R^\alpha f}(\vartheta_2)-f(2\alpha)|<\eta,\ \forall \vartheta_2\in f(2\alpha)+(-\bar\delta,\bar\delta).
\label{PRO1}
\end{equation}
By the property (P) above, let $\bar n_\delta$ sufficiently large so that 
\begin{equation}
\|F_{R^{n\tau}\theta_1^{(f)}(\vartheta_1,\cdot)}-F_{R^\alpha f}\|<\bar\delta-\eta,\ \forall n\geq \bar n_\delta.
\label{PRO2}
\end{equation}
Let $\bar k$ be sufficiently large so that $n_k\geq \bar n_\delta$ for all $k\geq \bar k$. 
Using the characterization of $\bar\delta$ above, let also $m_{\vartheta_2}\in\Z$ be such that 
\[
F_{f,\vartheta_1}^{(n_{\bar k})}(\vartheta_2)\in f(2\alpha)+m_{\vartheta_2}+(-\bar\delta,\bar\delta).
\]
Then by decomposing the next iterate as follows
\[
F_{f,\vartheta_1}^{(n_{\bar k}+1)}(\vartheta_2)=F_{R^{(n_{\bar k}+1)\tau}\theta_1^{(f)}(\vartheta_1,\cdot)}\circ F_{f,\vartheta_1}^{(n_{\bar k})}(\vartheta_2)-F_{R^\alpha f}\circ F_{f,\vartheta_1}^{(n_{\bar k})}(\vartheta_2)+F_{R^\alpha f}(F_{f,\vartheta_1}^{(n_{\bar k})}(\vartheta_2)),
\]
and using \eqref{PRO1} and \eqref{PRO2} and periodicity, we obtain 
\[
F_{f,\vartheta_1}^{(n_{\bar k}+1)}(\vartheta_2)\in f(2\alpha)+m_{\vartheta_2}+(-\bar\delta,\bar\delta),
\]
and then by induction
\[
F_{f,\vartheta_1}^{(n)}(\vartheta_2)\in f(2\alpha)+m_{\vartheta_2}+(-\bar\delta,\bar\delta),\ \forall n\geq n_{\bar k}.
\]
This proves that, under the assumption \eqref{BOUNDEDAWAY}, every accumulation point of the sequence $\{F_{f,\vartheta_1}^{(n)}(\vartheta_2)\}_{n\in\N}$ must lie in $f(2\alpha)+m_{\vartheta_2}+[-\bar\delta,\bar\delta]$. However, by continuity of $F_{R^\alpha f}$ and the property (P) above, the set $\text{Acc}(\{F_{f,\vartheta_1}^{(n)}(\vartheta_2)\}_{n\in\N})$ of these accumulation points must be invariant under $F_{R^\alpha f}$, {\sl viz.}\
\begin{equation}
F_{R^\alpha f}(\text{Acc}(\{F_{f,\vartheta_1}^{(n)}(\vartheta_2)\}_{n\in\N}))\subset \text{Acc}(\{F_{f,\vartheta_1}^{(n)}(\vartheta_2)\}_{n\in\N}).
\label{INVARACC}
\end{equation}
Since the only invariant set of $F_{R^\alpha f}$ that intersects $f(2\alpha)+m_{\vartheta_2}+[-\bar\delta,\bar\delta]$ is the fixed point $f(2\alpha)$, we conclude 
\[
\lim_{n\to+\infty}F_{f,\vartheta_1}^{(n)}(\vartheta_2)=f(2\alpha)+m_{\vartheta_2},
\]
as desired.

To conclude the proof, it remains to prove the existence of a unique $\vartheta_\text{unst,2}\in \T$ such that 
\[
\lim_{n\to+\infty}F_{f,\vartheta_1}^{(n)}(\vartheta_2)=f(t_\text{unst}+\alpha)\ (\text{mod}\ 1)\ \Longleftrightarrow\ \vartheta_2=\vartheta_\text{unst,2}.
\]
According to the first part of the proof, it suffices to prove the existence of a unique $\vartheta_\text{unst,2}$ such that the sequence $\{F_{f,\vartheta_1}^{(n)}(\vartheta_\text{unst,2})\}_{n\in\N}$ eventually enters and remains inside the interval $\text{I}$ where $F_{R^\alpha f}$ is expanding (because, henceforth, the sequence must approach $f(t_\text{unst}+\alpha)\ (\text{mod}\ 1)$). 

Let $\gamma>1$ be the derivative of $F_{R^\alpha f}$ on $\text{I}$. We claim (and prove below) the existence of a closed subinterval $\text{I}'\subset \text{I}$, of a number $\gamma'>1$ and of a sufficiently large $n'\in\N$ such that 
\begin{equation}
F_{R^{n\tau}\theta_1^{(f)}(\vartheta_1,\cdot)}(\text{I}')\supset \text{I}'\quad\text{and}\quad \left(F_{R^{n\tau}\theta_1^{(f)}(\vartheta_1,\cdot)}|_{\text{I'}}\right)'\geq\gamma',\ \forall n\geq n'.
\label{PROPAPPROX}
\end{equation}
More precisely, $\text{I}'$ is not unique and its boundaries can be chosen arbitrarily close to those of $\text{I}$ (and similarly $\gamma'$ can be chosen arbitrarily close to $\gamma$). Accordingly, the threshold $n'$ depends on $\text{I}'$ (and $\gamma'$) and diverges as $\text{I}'$ approaches $\text{I}$.

Now, given $n\geq n'$, consider the intersection set 
\[
\text{I}'\cap \bigcap_{k=n'}^nF^{-1}_{R^{n'\tau}\theta_1^{(f)}(\vartheta_1,\cdot)}\circ\cdots \circ F^{-1}_{R^{k\tau}\theta_1^{(f)}(\vartheta_1,\cdot)}(\text{I}').
\]
Thanks to the property \eqref{PROPAPPROX}, these sets form a family of nested closed non-empty intervals whose diameter tend to 0 as $n\to +\infty$. By the Nested Ball Theorem, we conclude the existence of a unique point $\bar\vartheta_2\in\text{I}'$ which depends on the whole solution $\theta_1$ at site 1, such that 
\[
F_{R^{n\tau}\theta_1^{(f)}(\vartheta_1,\cdot)}\circ\cdots \circ F_{R^{n'\tau}\theta_1^{(f)}(\vartheta_1,\cdot)}(\bar\vartheta_2)\in \text{I}',\ \forall n\geq n'.
\]
Furthermore, as a composition of homeomorphisms, the map $F^{(n'-1)}_{\theta_1}$ is itself a homeomorphism of $\T$. Hence there exists a unique $\vartheta_\text{unst,2}\in\T$ such that $F^{(n'-1)}_{\theta_1}(\vartheta_\text{unst,2})=\bar\vartheta_2$. Clearly, we have 
\[
F^{(n)}_{f,\vartheta_1}(\vartheta_\text{unst,2})\in \text{I}',\ \forall n\geq n',
\]
as desired. 

To complete the proof, it remains to establish the property \eqref{PROPAPPROX}. Notice first that, as $n$ gets large, the times at which the forcing signals $R^{n\tau}\theta_1^{(f)}(\vartheta_1,\cdot)$ and $R^\alpha f$ respectively cross the levels 0, $a_0$ and 1 come close together. As a consequence, the restriction of $F_{R^{n\tau}\theta_1^{(f)}(\vartheta_1,\cdot)}$ to the interval $\text{I}$, since close to $F_{R^{\alpha}f}|_{\text{I}}$ by the property (P), must consist of an expanding piece, possibly preceded and/or followed by a rigid rotation. 

Moreover, the length of these (putative) rigid rotation pieces must be bounded above by a number that depends only on $\gamma$ and on $\|F_{R^{n\tau}\theta_1^{(f)}(\vartheta_1,\cdot)}-F_{R^\alpha f}\|$, and which vanishes as $n\to +\infty$. (Indeed, if otherwise, the length(s) of the rigid rotation piece(s) remained bounded below by a positive number, we would have a contradiction with the property (P) above, because the distance between $F_{R^{n\tau}\theta_1^{(f)}(\vartheta_1,\cdot)}(\vartheta_2)$ and $F_{R^\alpha f}(\vartheta_2)$ would remain bounded below by a positive number for $\vartheta_2$ close to the boundaries of $\text{I}$.) This proves the existence of $\text{I}'\subset \text{I}$ on which all $F_{R^{n\tau}\theta_1}$ for $n$ sufficiently large, must be expanding.

In addition, by taking $n$ even larger if necessary (so that $\|F_{R^{n\tau}\theta_1^{(f)}(\vartheta_1,\cdot)}-F_{R^\alpha f}\|$ is even smaller), we can make sure that the expanding piece of $F_{R^{n\tau}\theta_1^{(f)}(\vartheta_1,\cdot)}$ over $\text{I}'$ intersects the diagonal, since the limit map $F_{R^\alpha f}$ does so. The first claim of property \eqref{PROPAPPROX} then easily follows. 

The second claim that the expanding slope must be bounded below by $\gamma'$ can proved using a similar contradiction argument as above. The proof of the Proposition is complete. 
\hfill $\Box$

In order to extend the stability results down the chain, proceeding similarly as when introducing $F_{f,\vartheta_1}^{(n)}$ before Proposition \ref{STAB2SITE}, we consider the maps $F_{f,\vartheta_1,\cdots,\vartheta_s}^{(n)}$ ($s>1$) that specify the coordinates $\theta_s^{(f)}(\vartheta_1,\cdots,\vartheta_s,n\tau)-n$. In particular, for $s=2$, a similar reasoning as the one in the proof of the property (P) above shows that the conclusion of Proposition \ref{STAB2SITE} implies 
\[
\lim_{n\to+\infty}\|F_{R^{n\tau}\theta_2^{(f)}(\vartheta_1,\vartheta_2,\cdot)}-F_{R^{2\alpha}f}\|=0,
\]
as soon as $\vartheta_1\neq\vartheta_\text{unst}$ and $\vartheta_2\neq \vartheta_\text{unst,2}$. Moreover, by repeating {\sl mutatis mutandis} the proof of Proposition \ref{STAB2SITE}, we conclude that, under the same conditions, there exists a unique $\vartheta_\text{unst,3}\in\T$ such that 
\[
\lim_{n\to+\infty}F_{f,\vartheta_1,\vartheta_2}^{(n)}(\vartheta_3)=f(3\alpha)\ (\text{\rm mod}\ 1),\ \forall \vartheta_3\neq \vartheta_\text{unst,3}.
\]
By induction, one obtains the following statement, from which Theorem \ref{MAINRES15} easily follows.
\begin{Cor}
Given an arbitrary $s\in\N$, assume that the initial phases $\vartheta_1,\cdots,\vartheta_s$ have been chosen so that the solution behaves as follows
\[
\lim_{t\to+\infty}\theta_{r}^{(f)}(\vartheta_1,\cdots,\vartheta_r,t)-f(t+\alpha r)=0\ (\text{\rm mod}\ 1), \forall r\in\{1,\cdots,s\}.
\]
Then there exists a unique $\vartheta_{unst,s+1}\in\T$ so that we have
\[
\lim_{n\to+\infty}F_{f,\vartheta_1,\cdots,\vartheta_s}^{(n)}(\vartheta_{s+1})=f(\alpha(s+1))\ (\text{\rm mod}\ 1),\ \forall \vartheta_{s+1}\neq \vartheta_{\text{unst},s+1}.
\]
\label{STABTW} 
\end{Cor}

\section{Generation of TW: proof of Theorem \ref{MAINRES2}}\label{S-GENER}
Recall that for $\tau\in I'_{\epsilon,a_0,a_1}$ we have $f(\sigma)<a_0$ for the associated TW where $\sigma=\frac{a_1-\alpha}{1+\epsilon}$ is such that $f(\alpha+\sigma)=a_1$.

Hence, for every continuous increasing and $\tau$-periodic forcing $g$ such that $g(0)=0$ and $g(\sigma)<a_0< g(\tau-\alpha)$, the stroboscopic map in the neighborhood of $f(\alpha)$ remains unchanged, {\sl viz.}\ 
\[
F_{g}(\vartheta_1)=F_f(\vartheta_1),\ \forall \vartheta_1\ \text{close to}\ f(\alpha)\ (\text{\rm mod}\ 1).
\]
By repeating {\sl mutatis mutandis} the reasoning in the proof of Proposition \ref{GLOBSTAB}, one shows that the restriction of $F_{g}$ to $[0,1)$ also consists of four pieces; two rigid rotations interspersed by a contracting and an expanding piece. Therefore, there must exist $\vartheta_{\text{unst},g}$ such that 
\[
\lim_{n\to+\infty}F_{g}^n(\vartheta_1)=f(\alpha)\ (\text{\rm mod}\ 1),\ \forall \vartheta_1\in\T\ :\ \vartheta_1\neq \vartheta_{\text{unst},g}.
\]
In other words, the asymptotic dynamics of $\theta_1$ is the same as when forcing with $f$. The conclusion for the rest of the chain $\{\theta_s\}_{s>1}$ immediately follows.

Finally, in order to make sure that the conclusion applies to the uniform forcing $g(t)=\frac{t}{\tau}$, it suffices to show that $\frac{\sigma}{\tau}<a_0<1-\frac{\alpha}{\tau}$. Using the expression $\tau=1-\epsilon\sigma$ and $\sigma=\frac{a_1-\alpha}{1+\epsilon}$, we obtain after simple algebra
\begin{equation}
\frac{a_1-a_0-\epsilon a_0(1-a_1)}{1+\epsilon a_0}<\alpha<\frac{(1-a_0)(1+\epsilon(1-a_1))}{1+\epsilon a_0}.
\label{STRICTFINAL}
\end{equation}
It is easy to see that this condition defines an interval of $\alpha$ for all parameter values $\epsilon, a_0,a_1$. Moreover, and depending whether $a_1$ is smaller than $1-a_0$ or not, for $\epsilon=0$, this interval either contains or coincides with the interval defined in the proof of Corollary \ref{GLOBEXISTS}. 
By continuity, for $\epsilon>0$ sufficiently small, there exists an interval of $\alpha$ - that corresponds to periods in a subinterval $I''_{\epsilon,a_0,a_1}\subset I'_{\epsilon,a_0,a_1}$ - in which both the condition \eqref{STRICTFINAL} and the ones in the proof of Corollary \ref{GLOBEXISTS} hold. The statement easily follows.

\section{Concluding remarks}\label{S-CONCL}
For simple feed forward chains of type-I oscillators, our analysis proved that periodic wave trains can be generated from arbitrary initial condition, even when the root node is forced using an unrelated signal. Moreover, these stable waves exist for an open (parameter-dependent) interval of wave number and period. 

The existence of globally attracting waves for arbitrary wave number in some range is reminiscent of the inertia-free dynamics of tilted Frenkel-Kontorova chains, which constitute coupled oscillator models for spatially modulated structures in solid-state physics \cite{FM96}.  There are however essential differences between the two situations. Instead of a uni-directional interaction, the coupling is of bi-directional type in Frenkel-Kontorova chains and involves left and right neighbors. More importantly, the overall dynamics there is monotonic and, as mentioned in the introduction, this property is critical for the proof of existence and stability of waves \cite{BM98}.

Finally, we notice that the results on asymptotic stability and on stability with respect to changes in forcing are based on hyperbolic properties of the stroboscopic dynamics. Accordingly, we believe that, using continuation methods, these results, and more generally, results on generation of traveling waves in unidirectional chains of type-I oscillators can be established in a rigorous mathematical way, in more general models with smooth PRC and stimulus nonlinearities. This will be the subject of future studies.

\section*{Competing interests}
The authors declare that they have no competing interests.
  
\section*{Author's contributions}
BF and SM both designed the research, proceeded to the analysis, and wrote the paper.

\section*{Acknowledgements} 

The work of BF was supported by EU Marie Curie fellowship PIOF-GA-2009-235741 and by CNRS PEPS {\em Physique Th\'eorique et ses interfaces}.

\end{document}